\documentclass[aps,groupedaddress,twocolumn,floatfix,showpacs,tightenlines]{revtex4-1}
\usepackage{graphicx}
\usepackage{float}
\usepackage{color}
\usepackage{psfrag}
\usepackage{dcolumn}
\usepackage{bm}

\usepackage{amssymb}
\usepackage{amsmath}
\usepackage{amsfonts}
\usepackage{longtable}
\usepackage{xspace}
\usepackage{booktabs}
\usepackage{mathrsfs}
\usepackage{gensymb}
\usepackage[table]{xcolor}

\newcommand{\beq}{\begin{equation}}
\newcommand{\eeq}{\end{equation}}
\newcommand{\bea}{\begin{eqnarray}}
\newcommand{\eea}{\end{eqnarray}}
\newcommand{\bes}{\begin{subequations}}
\newcommand{\ees}{\end{subequations}}

\newcommand{\scri}{\mathscr{I}}

\begin{document}
\title{Adapted gauge to small mass ratio binary black hole evolutions}
\author{Nicole Rosato, James Healy, and Carlos~O.~Lousto}
\affiliation{Center for Computational Relativity and Gravitation,\\
School of Mathematical Sciences,
Rochester Institute of Technology, 85 Lomb Memorial Drive, Rochester,
New York 14623}
\date{\today}

\begin{abstract}
We explore the benefits of adapted gauges to small mass ratio binary black hole
evolutions in the moving puncture formulation. We find expressions that
approximate the late time behavior of the lapse and shift, 
$(\alpha_0,\beta_0)$, and use them as
initial values for their evolutions. We also use a position and black hole
mass dependent damping term, $\eta[\vec{x}_1(t),\vec{x}_2(t),m_1,m_2]$, 
in the shift evolution, rather than a constant 
or conformal-factor dependent choice. 
We have found that this substantially
reduces noise generation at the start of the numerical integration
and keeps the numerical grid stable around both black holes, allowing for more accuracy with
lower resolutions.
We test our choices for this gauge in detail in a case study of a binary with a
7:1 mass ratio, and then use 15:1 and 32:1 binaries for a convergence study. Finally, we apply our new gauge to a 64:1 binary and a 128:1 binary to well cover 
the comparable and small mass ratio regimes.
\end{abstract}
\pacs{04.25.dg, 04.25.Nx, 04.30.Db, 04.70.Bw}

\maketitle


\section{Introduction}\label{sec:intro}

The 2005 breakthrough in numerical relativity techniques 
\cite{Pretorius:2005gq,Campanelli:2005dd,Baker:2005vv} has
allowed for the production of thousands of binary black hole simulations 
(see for instance the catalogs 
\cite{Boyle:2019kee,Jani:2016wkt,Healy:2019jyf,Healy:2020vre}). 
Improvements in hardware and numerical techniques contribute 
by speeding up simulations, however, some corners of the binary's parameter space 
remain relatively unexplored.
Binaries in the high mass-ratio regime are challenging to simulate; they 
can take months of supercomputer time and require substantial computational resources to run accurate, long term, numerical evolutions. These types of binaries are of particular interest for third generation gravitational wave
detectors \cite{Purrer:2019jcp} and for the space-based mission LISA \cite{Gair:2017ynp} since they have long, low frequency inspiral periods.
Prototype small mass ratio simulations reaching 100:1 have been achieved with
the moving puncture approach \cite{Lousto:2010ut,Sperhake:2011ik} 
and numerical convergence has been proven.
Recently, another sequence of nonspinning binaries with mass ratios
$q=m_1/m_2=1/32, 1/64, 1/128$ has been studied in \cite{Lousto:2020tnb}. 
Such simulations should be considered proof of principle, but in order to become
practical for production purposes, they need improvements in both computational efficiency
and accuracy of the numerical techniques. Of particular interest in this
paper is to explore the choice for the numerical gauge, as well as the initial values for the lapse and shift equations,
as a means to achieve improvements in accuracy without requiring
more highly resolved (and hence more expensive) simulations.

In 2005, a fundamental breakthrough was the choice of the gauge equations. The gauge was originally developed to force successful evolutions without numerical simulations crashing.  The moving puncture 
approach proved robust and produced accurate waveforms,
even allowing the evolution of multi-black hole systems \cite{Lousto:2007rj}.
However, when studied in detail, subtleties appear with the convergence
\cite{Zlochower:2012fk}, and some gauge amplified initial noise has 
been observed \cite{Etienne:2014tia}.
This is particularly relevant for binaries with small mass ratios ($q=m_1/m_2$),
as the amplitude of gravitational radiation scales like $\sim q$. The
initial noise then reflects at the boundaries of the mesh refinement levels, which are
necessary to efficiently describe the different scales of the binary system, causing high frequency oscillations when those reflections reach the observers.

In this paper we will show that choices of the initial lapse and a shift damping parameter $\eta$
in the gauge 
can cure those initial inaccuracies,
and lead to a much cleaner evolution of the binary black holes. We explore different choices of the initial lapse and shift
in Section \ref{sec:gauges} to improve the accuracy of the simulations. In a previous paper \cite{Healy:2020iuc} we studied
the effect of different, constant shift damping parameters $\eta$ on the
extraction of recoil velocities from the horizon of the final remnant
black hole. In this paper, in Section \ref{sec:smallq}
we extend the analysis to
adapt $\eta(\vec{x}_1(t),\vec{x}_2(t))$ to small mass ratio binaries.
The results using these numerical techniques on the simulation of a prototypical nonspinning binary with
mass ratio $q=1/7$ is studied in detail in Section \ref{sec:7results}
with different choices
for $\eta$ and the initial lapse, as well as control convergence studies
of binaries with mass ratio $q=1/15$ and then $q=1/32$ in Sections \ref{sec:15results} and \ref{sec:32results} respectively. In Section \ref{sec:q64128results}, we provide results for the extremal $q=1/64$ binary using the gauge choice we determined from the previous sections.

We conclude in Section \ref{sec:discussion}
with an optimal selection of initial lapse and shift damping parameter $\eta$ that
is simple to implement and numerically efficient, while still improving the accuracy
of the simulations in the small mass ratio regime.

\section{Numerical Techniques}\label{sec:nr}
The 2005 breakthrough work \cite{Campanelli:2005dd} has allowed us to 
obtain accurate, convergent waveforms and horizon parameters by
evolving the BSSNOK \cite{Nakamura87, Shibata95, Baumgarte99}
system in conjunction with a modified 1+log lapse and a
modified Gamma-driver shift condition~\cite{Alcubierre02a,Campanelli:2005dd},
  \begin{eqnarray}
\partial_0\alpha=(\partial_t - \beta^i \partial_i) \alpha &=& - 2 \alpha K,\\
 \partial_t \beta^a &=& \frac34 \tilde \Gamma^a - \eta(x^k,t)\, \beta^a.\label{eq:gauge}
  \end{eqnarray}
with the initial shift vanishing and the initial lapse $\alpha_0= 2/(1+\psi_0^{4})$,
where the conformal factor is defined as
\beq\label{eq:conformal}
\psi_0=1+\frac{m_1}{2|\vec{r}-\vec{r}_1|}+\frac{m_2}{2|\vec{r}-\vec{r}_2|}. 
\eeq
Here and in the remainder of this paper, Latin indices 
such as $i\text{ and }k$ cover the spatial range $1,2,3$. Our units use the $G=c=1$ convention.

The parameter $\eta$ (dimensions one-over-mass: $1/m$) 
in the shift equation regulates the damping of the gauge oscillations.
We have found in \cite{Lousto:2007db} that coordinate dependent measurements, 
such as spin and linear momentum direction, become more accurate as $\eta$ is 
reduced and the grid resolution is extrapolated to infinite ($h\to0$). 
However, if $\eta$ becomes too small $(\eta\ll1/m)$, 
the runs may become unstable. Similarly, if $\eta$ is too large $(\eta\gg10/m)$, 
then grid stretching effects can cause the remnant horizon to continuously grow, 
eventually leading to an unacceptable loss in accuracy at late times. 
Therefore, $\eta$ is commonly chosen to be of order unity as a compromise 
between the accuracy and stability of binary black hole evolutions; for comparable-mass binaries, our standard choice is
$\eta = 2/m$.

To compute the initial data for the lapse and shift equations, we use
the {\sc TwoPunctures}~\cite{Ansorg:2004ds} thorn.
These black-hole-binary (BHB) data-sets are then evolved using the
{\sc LazEv}~\cite{Zlochower:2005bj} implementation of the moving
puncture formalism~\cite{Campanelli:2005dd}.
The Carpet~\cite{Schnetter-etal-03b,carpet_web} mesh refinement
driver provides a `moving boxes' style mesh refinement and
we use {\sc AHFinderDirect}~\cite{Thornburg2003:AH-finding} to locate
apparent horizons.
The magnitude of the horizon mass, spin, and linear momentum are computed using
the {\it isolated horizon} (IH) algorithm detailed in
Ref.~\cite{Dreyer02a} (as  implemented in
Ref.~\cite{Campanelli:2006fy}).
Once we have the horizon spin, we can calculate the horizon
mass via the Christodoulou formula 
${m_H} = \sqrt{m_{\rm irr}^2 + S_H^2/(4 m_{\rm irr}^2)}\,,$
where $m_{\rm irr} = \sqrt{A/(16 \pi)}$ and  $A$ is the surface area
of the horizon. 
The radiated energy, linear momentum, and angular momentum are all measured in
terms of the Newman-Penrose Weyl scalar $\Psi_4$, using the formulae provided in
Refs.~\cite{Campanelli99,Lousto:2007mh}, and extrapolation to to $\scri^+$
is performed with the formulas given in Refs.~\cite{Healy:2020iuc}~\cite{Nakano:2015pta}~\cite{Krishnan:2007pu}.

Convergence studies of our simulations have been performed
in Appendix A of Ref.~\cite{Healy:2014yta},
in Appendix B of Ref.~\cite{Healy:2016lce}, and
for nonspinning binaries are reported in Ref.~\cite{Healy:2017mvh}.
For very highly spinning black holes ($s/m^2=0.99$)
convergence of evolutions was studied in Ref. \cite{Zlochower:2017bbg},
for precessing $s/m^2=0.97$ in Ref. \cite{Lousto:2019lyf}, and
for ($s/m^2=0.95$) in Ref. \cite{Healy:2017vuz} for unequal mass binaries.
These studies allow us to assess that the simulations presented here,
with similar grid structures, are well
resolved by the adopted resolutions and are in a convergence regime.

\subsection{The Initial Gauge}\label{sec:gauges}

In this section we derive the form of a new set of equations for the initial lapse and shift. 
The goal is to approximate those of the trumpet slice in quasi-isotropic 
coordinates $r$, both near the
puncture $r=0$ and far from source as powers of $1/r$. We will do this
for a single black hole, then we will superpose the result for two black holes.


\subsubsection{Initial Lapse}


To construct the trumpet-like Late Time initial Lapse (LTL), we begin by proposing the following form:
\beq\label{eq:LTL}
\alpha_{LTL}=\alpha_0(\psi_0)=\frac{a}{1+b\psi_0^n+c\psi_0^{n+1}+d\psi_0^{n-1}}
\eeq
where $a,b,c$ and $d$ are constants to be determined by matching to 
trumpet data in isotropic coordinates close to the punctures, and to the behavior of the lapse far away $\alpha\sim(1-m/r)/(1+m/r)$. The value $n$ is a function of an unknown constant $\gamma$, which is to be determined later.

The
initial lapse is written as a function of the conformal factor $\psi_0$ defined in Eq.~(\ref{eq:conformal})
as an extension of our original standard form for the lapse
\beq\label{eq:initlapse}
\alpha_0= 2/(1+\psi_0^{4}).
\eeq
We have the option of superposing the individual initial lapses for 
each puncture, so that 
$\alpha_i=\sum_{i=1,2}\alpha_{LTL}(\psi_i)$ where 
\beq\label{eq:iconformal}
\psi_i = 1 +\frac{m_i}{2|\vec{r}-\vec{r}_i|},
\eeq
which would make our new initial lapse 
\beq
\alpha_{LTL}=\alpha_1+\alpha_2 -1
\eeq
but this leads to negative values near the punctures.

To obtain the desired behavior in Eq.~(\ref{eq:LTL}), we begin by expanding the Schwarzschild lapse in isotropic coordinates in powers of $\frac{1}{r}$ to obtain
\beq\label{Sch}
\alpha_{Sch}=(1-m/2r)/(1+m/2r)=1-1/r+1/2r^2+{\cal O}(1/r^3).
\eeq
Near the puncture, the expected trumpet-like behavior is 
\beq
\alpha\sim Ar^{1/\gamma}
\eeq
where (by Eq. (48) of \cite{Brugmann:2009gc})
\bea
\gamma&=&(2-R_0)/(6-4R_0)=0.9163407461;\\
 R_0&=&1.312408290
\eea
and according to numerical computations of
\cite{Thierfelder:2010dv}, $A=0.54$.
Using this value will allow us to
to match the numerical behavior rather than 
strictly the isotropic coordinates (as in \cite{Brugmann:2009gc}).

Choosing $n+1=1/\gamma$ in Eq.~(\ref{eq:LTL}) above and
taking the limit as $r\to0$, we obtain that
\beq
2.131254761(c/a)=0.54
\eeq
which is approximated to be $a/c=1/4$.

Setting the other constants $b,c,d$ to match the three orders
of the expansion in Eq. (\ref{Sch}), we find
\bea
a&=&{\frac {2\,\gamma-1}{6\,\gamma-1}},b=-10\,{\frac {2\,\gamma
-1}{6\,\gamma-1}},\\
c&=&4\,{\frac {2\,\gamma-1}{6\,\gamma-1}},d=2\,{\frac 
{4\,\gamma-3}{6\,\gamma-1}},\\n&=&-{\frac {\gamma-1}{\gamma}}\nonumber
\eea
These expressions are finally inserted into Eq.~(\ref{eq:LTL}) to construct this new choice for initial lapse.

For a visual representation of the differences between the typical initial lapse $\alpha_0= 2/(1+\psi_0^{4})$ and $\alpha_{LTL}$, refer to Fig.~\ref{fig:lapse}, which shows the two choices for lapse in blue and red (respectively) and their effects on a $q=1/3$ binary with initial separation $d=8m$. Here, we can see that $\alpha_{LTL}$ (in red) is tighter around the punctures than its counterpart in blue $\alpha_0$, therefore mimicking the shape of the settled lapse more accurately. 

\subsubsection{Initial Shift}

We would also like to find a formula to model the shift $\beta$ late-time behavior
analogous to Eq.~(\ref{eq:LTL}).  From \cite{Brugmann:2009gc} [Eqs. (7) and (18)]
 we have analytic expressions for the shift at distances close to the black hole 
\beq\label{shift:1}
\beta^r = r \beta/R
\eeq
and at large $r$, the shift magnitude, 
\beq \label{shift:2}
\beta^2 = C\exp(\alpha)/R^6,
\eeq
where $r$ denotes isotropic coordinates and $R$ denotes Schwarzschild ones and, 
$C=1.554309591$.
Equating the equations (\ref{shift:1}) and (\ref{shift:2}),
\bea\label{shift}
\beta^r/r&&=\sqrt(C)\exp(\alpha/2)/R^3\\
&&\to\sqrt(C)/R_0^3=0.5515207650,
\eea
for $R\to0$
where from Eqs. (23) and (28)  of \cite{Brugmann:2009gc} we have
\bea
C&&=e^{3-\sqrt{10}}(3+\sqrt(10)^3/128\nonumber\\
&&=1.554309591R_0^4-2R_0^3+C=0,\\
R_0&&=1.312408290
\eea

This agrees well with the estimates derived from the numerical fittings
(6) and (7) in \cite{Thierfelder:2010dv}. From these, define
\beq
K\approx0.30-0.92\alpha;\quad{\rm where}\quad K=\beta \alpha'(R)/2.
\eeq
From which we can find that the leading order term of the shift should be
\bea\label{shift:3}
\beta^r/r&&=\beta/R=2K/R\alpha'(R)\\
&&\to0.60/R_0\alpha'(R_0)\approx0.55,
\eea
for $r\to 0$.
With the use of Eq. (31) in \cite{Brugmann:2009gc} we have
\beq
\alpha'(R_0)=(6-4R_0)/(2-R_0)R_0.
\eeq

On the other hand, to study the shift behavior at large distances from
the BH we have in the expansion of the eq. (\ref{shift})
\beq\label{shift:4}
{\frac {\sqrt {C}{{\rm e}^{1/2}}}{{r}^{2}}}-7/2\,{\frac {\sqrt {C}{
{\rm e}^{1/2}}}{{r}^{3}}}+{\frac {57}{8}}\,{\frac {\sqrt {C}{{\rm e}^{
1/2}}}{{r}^{4}}}+O \left( {r}^{-5} \right) 
\eeq
where we have used that in isotropic coordinates,
\beq
\alpha_{iso}=(1-m/2r)/(1+m/2r) 
\eeq
far from the center of coordinates.

In a similar fashion as for the lapse, we propose the following representation for the initial shift

\beq\label{beta0}
\beta_0^r(\psi_0)=\frac{a (\psi_0-1)^2}{1+b\psi_0+c\psi_0^2+d\psi_0^3}
\eeq 

Matching expansions (\ref{shift:3}) and (\ref{shift:4}) with (\ref{beta0}), we get

\bea
a &= -0.5368350604, b = -3.620281004,\nonumber \\
c &= 4.501732957, d = -1.946744690.
\eea

\subsubsection{Two black holes}

It is our ultimate goal to use these analytic approximants as initial data for 
the gauge to evolve a binary system. 

%
\begin{figure}[ht]
  \includegraphics[angle=0,width=0.8\columnwidth]{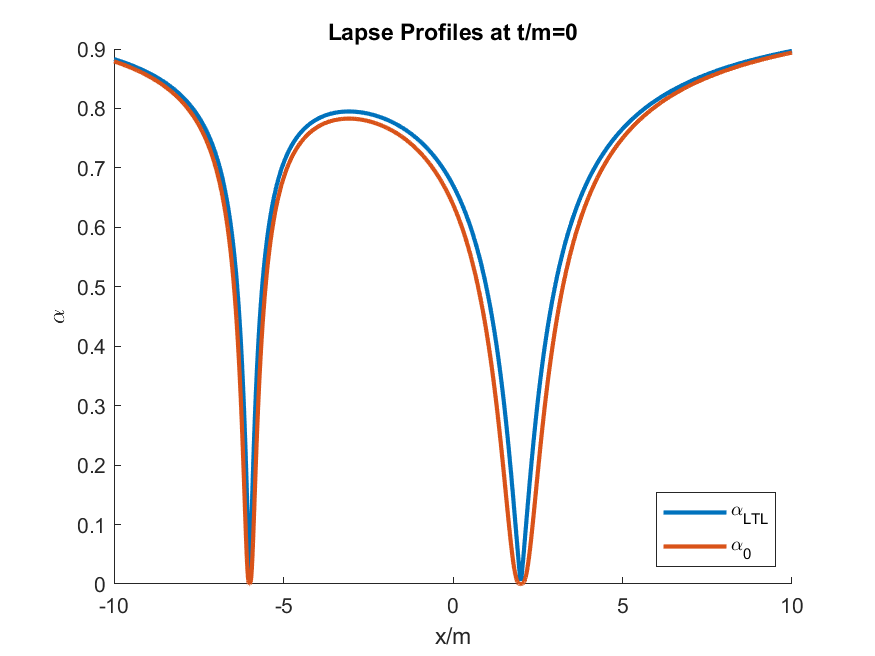}
  \caption{In red is shown the initial lapse $\alpha_0(\psi_0)$ and in blue is the late time lapse $\alpha_{LTL}$. [This is for a case study with $m_2=3/4$, $m_1=1/4$, located at $x_1=-6$ and $x_2=2$ respectively, and where we have normalized so that $m_1+m_2=1$].
\label{fig:lapse}}
\end{figure}

We do not assume that the lapse for each black hole adds linearly, but instead include the information about the binary in the conformal factor $\psi_0$. For an arbitrary binary, the $\psi_0$ in Eq.~(\ref{eq:LTL}) matches that in Eq.~(\ref{eq:conformal}).
The resulting shape of the initial lapse is shown in Fig.~\ref{fig:lapse} for a binary with $m_1=1/4$, $m_2=3/4$, located at $x_1=+6$ and $x_2=-2$, respectively and normalized by $m_1+m_2=1$.

\begin{figure}[h!]
  \includegraphics[angle=0,width=0.8\columnwidth]{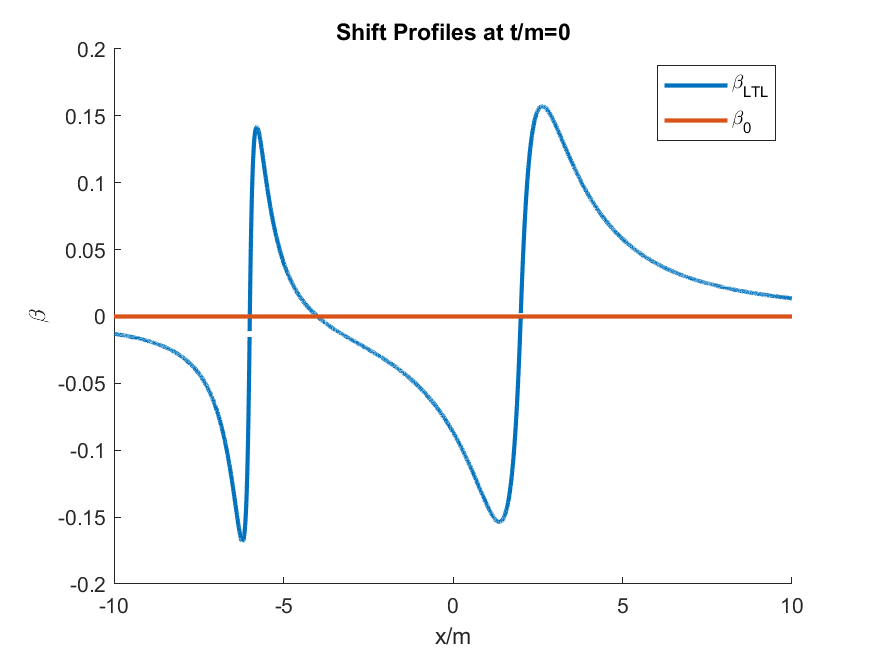}
  \caption{In red is the usual zero initial shift, and in blue is the superposition of the individual shifts showing the push away from each black hole. [This is for a case study with $m_2=3/4$, $m_1=1/4$, located at $x_1=2$ and $x_2=-6$, respectively].
\label{fig:shift}}
\end{figure}

In the case of the trumpet Late Time initial Shift (LTS), we do in fact assume it adds linearly for the two black holes as this matches the settled shape of the evolved shift best: 
\beq
\beta_{LTS}=\beta^r_1(\vec{r}-\vec{r}_1)/|\vec{r}-\vec{r}_1|+
\beta^r_2(\vec{r}-\vec{r}_2)/|\vec{r}-\vec{r}_2|.
\eeq
Fig.~\ref{fig:shift} displays the behaviors of the initial shifts (zero-shift in blue, and LTS in red) for the same case study as above. The LTS initial data pushes the shift away from the black holes at the punctures and damps to zero far away.

We have chosen to not superpose the lapse and to superpose the shifts because, while testing different configurations, we found that those choices best matched the late-time behavior of the lapse and shift.
The construction presented in this section ignores the motion of the black holes. The inclusion of initial linear momentum and spins of the holes into the analytic expressions for the initial lapse and shift can be done in terms of a Lorentz boost and a Kerr shift. We provide explicit expressions in the Appendix \ref{sec:LBKS}, but do not provide an in-depth study here because we found that their inclusion into the initial choice for the gauge is not crucial in the case of a nonspinning binary.
\subsection{Shift damping parameter $\eta$}\label{sec:smallq}

The principal purpose of this study is to investigate whether the accuracy of
small mass ratio binaries can be improved by modifications to the gauge equations Eq.~(\ref{eq:gauge}).  In this section, we seek to develop a superposed Gaussian model for 
the shift damping parameter $\eta$.
For comparable mass $q>1/10$ binary evolutions, our simulations typically use a constant 
$\eta$; in general we choose $\eta=2/m$, but recent studies have shown that $\eta=1/m$ may provide a better measure of recoil velocity at the horizon of the remnant black hole \cite{Krishnan:2007pu} (as in \cite{Healy:2020iuc}, in which we studied binaries as small as $q=1/5$). 

For mass ratios smaller that $q=1/10$, a non-constant $\eta$ is required for simulation stability, especially at lower grid resolutions. Ref.~\cite{Lousto:2010ut} introduced
$\eta(W)$ ($W=\sqrt{\chi}=\exp(-2\phi)$ where $\phi=\phi(\frac{1}{\psi_0})$, with the conformal factor
suggested by~\cite{Marronetti:2007wz}),
or modified (as below in equations (\ref{eq:conformal})-(\ref{eq:etaG})). The modification we use is
based on the superposition of weighted Gaussians with peaks at 
the punctures. Alternatives for the conformal factor 
have been suggested by Refs.
\cite{Mueller:2009jx,Lousto:2010tb,Muller:2010zze,Alic:2010wu,Schnetter:2010cz}, but they were not investigated
with respect to small-mass ratio binaries.

\begin{figure}[h!]
\includegraphics[angle=0,width=0.7\columnwidth]{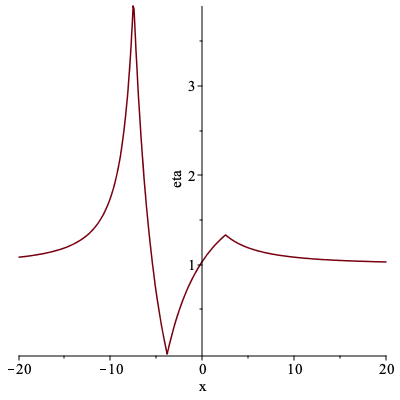}
\caption{$\eta_\psi$ profile for ($m=m_1+m_2=1$ here)
$m_2=3/4$, $m_1=1/4$; $x_1= 2.5$, $x_2=-7.5$; $a=1$, $b=2$; $A=1$, $B=1$. Note that this is technically different from the $\eta(W)$ used in \cite{Lousto:2010ut} since we use the specific form ($\ref{eq:conformal}$) for $\psi_0$ instead of the evolved variable, which is related to the inverse of the conformal factor $W=\sqrt{\chi}$ \cite{Campanelli:2005dd}.
\label{fig:eta0}}
\end{figure}

Here we bring back some of those ideas, where we evaluate
 $\eta(\vec{r}_1(t),\vec{r}_2(t))$ parameterized by the black 
holes' punctures trajectories $(\vec{r}_1(t),\vec{r}_2(t))$.
The (initial form of the) conformal factor evaluated at every time step is given by Eq. (\ref{eq:conformal})
and we can define, analogously to $\eta(W)$,
\beq\label{eq:eta0}
m\eta_\psi=\mathcal{A}+ \mathcal{B}\,{\frac {\sqrt {|\vec{\nabla}_r\psi_0|^2}}{ \left( 1-{\psi_0}^{a} \right) ^{b}}}\,.
\eeq
An example of this $\eta_\psi$ is plotted in Fig.~\ref{fig:eta0}. 
At the $i^{th}$-puncture $m_i\eta=1$ and at the center of mass $m\eta=1$, 
but Eq. (\ref{eq:eta0}) goes through a minimum $m\eta=0$ in between the holes
and at this point, as well as at the punctures, $\eta$ is $C^0$. 
Since the gauge condition Eq.~(\ref{eq:gauge}) involves an integration not a derivative, 
this does not affect numerical evolutions.

\begin{figure}[h!]
\includegraphics[angle=0,width=0.85\columnwidth]{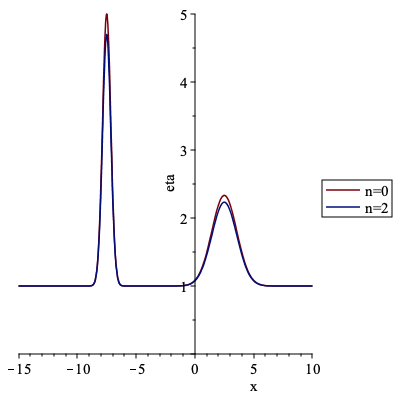}
\caption{$\eta_G$ profile for ($m=m_1+m_2=1$ here)
  $m_2=3/4$, $m_1=1/4$; $x_1= 2.5$, $x_2=-7.5$; $A=1$, $B=1$, $C=1$;
  $\sigma_1=2\,m_1$, $\sigma_2=2\,m_2$. At large separations, $n=0$ and $n=2$ show good agreement.
\label{fig:etaG}}
\end{figure}

A second alternative for smoother behavior is the superposition of Gaussian (See similarly \cite{Muller:2010zze})
\bea\label{eq:etaG}
\eta_G=&&\frac{\mathcal{A}}{m}+\frac {\mathcal{B}}{m_1}\left(\frac{\vec{r}_1(t)^2}{\vec{r}_1(t)^2+\sigma_2^2}\right)^n
e^{-\left|\vec{r}-\vec{r}_1(t)\right|^{2}/\sigma_1^{2}}\nonumber\\
&&+\frac {\mathcal{C}}{m_2}\left(\frac{\vec{r}_2(t)^2}{\vec{r}_2(t)^2+\sigma_1^2}\right)^n
e^{-\left|\vec{r}-\vec{r}_2(t)\right|^{2}/\sigma_2^{2}},
\eea
which, for the punctures of the previous example, is displayed 
in fig. \ref{fig:etaG} (with $n=0$), and behaves like $m_2\eta_G=1.25$ at the first puncture,
and as $m_1\eta_G=1.75$ at the second puncture, and it goes to 1
in between and far away from the binary.

\begin{figure}[ht]
\includegraphics[angle=0,width=0.85\columnwidth]{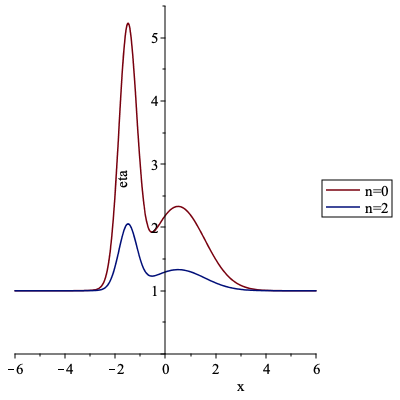}
\caption{$\eta_G$ profile for ($m=m_1+m_2=1$ here)
  $m_2=3/4$, $m_1=1/4$; $x_1= 0.5$, $x_2=-1.5$; $A=1$, $B=1$, $C=1$;
  $\sigma_1=2\,m_1$, $\sigma_2=2\,m_2$. At small separations $n=2$ reduces the peaks in Eq.~(\ref{eq:etaG}) that using $n=0$ is designed to produce.
\label{fig:etaGC}}
\end{figure}

We have applied Eq.~(\ref{eq:etaG}) (labelled $n=0$) to simulations of small-mass ratio binaries (as $\eta_G$ or G), and compared the physical output to both the constant $\eta=2/m$ as well as the $\eta(W)$ gauges. The results are forthcoming in the following sections.
The introduction of the $n=2$ case is to model smaller effective spikes once the black holes are merged as shown in Fig.~\ref{fig:etaGC}. For larger separation $n=0$ and $n=2$ essentially agree with each other as shown in Fig. \ref{fig:etaG}.

\section{Simulation Results}\label{sec:results}
  
Here we present the results of our simulations using different choices of initial lapse and choices for $\eta$. We proceed in a descendent mass ratio, from the $q=1/7$ and $q=1/15$, and then onto the more challenging $q=1/32$ to find the best gauge choices and apply them to the most challenging case, with mass ratio $q=1/64$. For the rest of this work, the following notation will be used: `LZ' denotes the Lousto-Zlochower $\eta=\eta(W)$ gauge, `2' denotes the constant $\eta=2/m$, `G' denotes the Gaussian in Eq.~(\ref{eq:etaG}) with $n=0$, and the `+ LTL' denotes the addition of the Late Time Lapse choice for initial data. Table \ref{tab:simlist} shows a full list of all the simulations performed for this paper and their initial parameters. All simulations use 8th order finite differencing in space.

The resolutions in Table \ref{tab:simlist} are listed in the form nXXX where XXX is the number of gridpoints on the coarsest grid level (i.e. n100 has 100 points on the coarsest grid level). The mesh increases by a factor of two in resolution per refinement level, with the most refined levels surrounding the individual punctures.

\begin{table*}[t]
\begin{tabular}{|llllllllll|}\hline
     Run&q&D&x1/m&x2/m&$P_r/m$&$P_t/m$&$m\eta$&CFL&Resolution  \\\hline\hline
     1&1/7&11m&1.375&-9.625&-1.42e-4&0.0396&LZ &1/3& n100 \\  
     2&1/7&11m&1.375&-9.625&-1.42e-4&0.0396&2/m &1/3& n100 \\  
     3&1/7&11m&1.375&-9.625&-1.42e-4&0.0396&G &1/3& n100 \\  
     4&1/7&11m&1.375&-9.625&-1.42e-4&0.0396&LTL+G &1/3& n100 \\  \hline
     5-7&1/15&8.5m&0.5378&-7.9622&-1.04e-4&0.0256&LZ &1/3& n084,n100,n120\\ 
     8&1/15&8.5m&0.5378&-7.9622&-1.04e-4&0.0256&2/m &1/4& n084\\ 
     9,10&1/15&8.5m&0.5378&-7.9622&-1.04e-4&0.0256&2/m &1/3& n100,n120\\ 
     11-13&1/15&8.5m&0.5378&-7.9622&-1.04e-4&0.0256&G& 1/3&n084,n100,n120\\\hline
     14-16&1/32&8.00m&0.2424&-7.7576&-3.32e-5&0.0135&LZ &1/4& n084,n100,n120\\ 
     17&1/32&8.00m&0.2424&-7.7576&-3.32e-5&0.0135&2/m &1/4& n100\\ 
     18-20&1/32&8.00m&0.2424&-7.7576&-3.32e-5&0.0135&G& 1/4&n084,n100,n120\\\hline
     21,22&1/64&7.00m&0.1077&-6.8923&-1.49e-5&0.0077&LZ &1/4& n084,n100\\ 
     23&1/64&7.00m&0.1077&-6.8923&-1.49e-5&0.0077&G& 1/4&n100\\\hline
     24&1/128&7.00m&0.0543&-6.9457&-3.85e-6&0.0039&LZ& 1/4&n100\\
     25&1/128&7.00m&0.0543&-6.9457&-3.85e-6&0.0039&G& 1/4&n100 \\\hline
\end{tabular}
\caption{A full of initial data parameters for the simulations performed. The smaller black hole is labeled 2 and larger black hole labeled 1. The punctures are located at $r_i=(x_i,0,0)$ with initial momenta $P=\pm(P_r,P_t,0)$ and mass-ratio q. All simulations are nonspinning. A full study with 3 resolutions for all gauge choices was done on the $q=1/15$ binary, but results are shown for simulations 5-7, 9, and 11-13 because simulation 8 could only be completed with a in increase in time resolution (CFL $1/3\to1/4$). The $q=1/32$ binary has three resolutions for the LZ and G gauge choices. The smaller $q=1/64$ and $q=1/128$ as well as the larger $q=1/7$ mass ratios are used for verification. All simulations use eighth order finite differencing in space. }
\label{tab:simlist}
\end{table*}

\subsection{Results for a $q=1/7$ nonspinning binary}\label{sec:7results}
In this section we will begin our analysis by studying the effects of different gauge modifications on
the physical parameters of a binary system. We will first verify it works on a comparable mass binary, with mass ratio $q=m_1/m_2=1/7$ and binary separation $D=11m$.
This system, while not as computationally intensive as the smaller mass ratio binaries studied later on, is still fairly nontrivial, 
and will serve to help generalize our results for the other mass-ratio systems. 

For this system we did four runs with four different choices for the gauge all at our typical production resolution n100. The first uses the choice LZ, and does not modify initial lapse and shift from the standard $\alpha_0 = 2/(1+\psi_0^4)$ and $\beta_0 = 0$. This is our reference choice for $\eta$ for small mass ratio runs. The second simulation uses G, the third run uses the constant choice $\eta=2/m$ and, finally, the fourth run uses G + LTL.
All use eighth order spatial finite differencing stencils and  fourth order Runge-Kutta in time, with a Courant Factor of 1/3.



\begin{figure}[h!]
\includegraphics[angle=0,width=0.99\columnwidth]{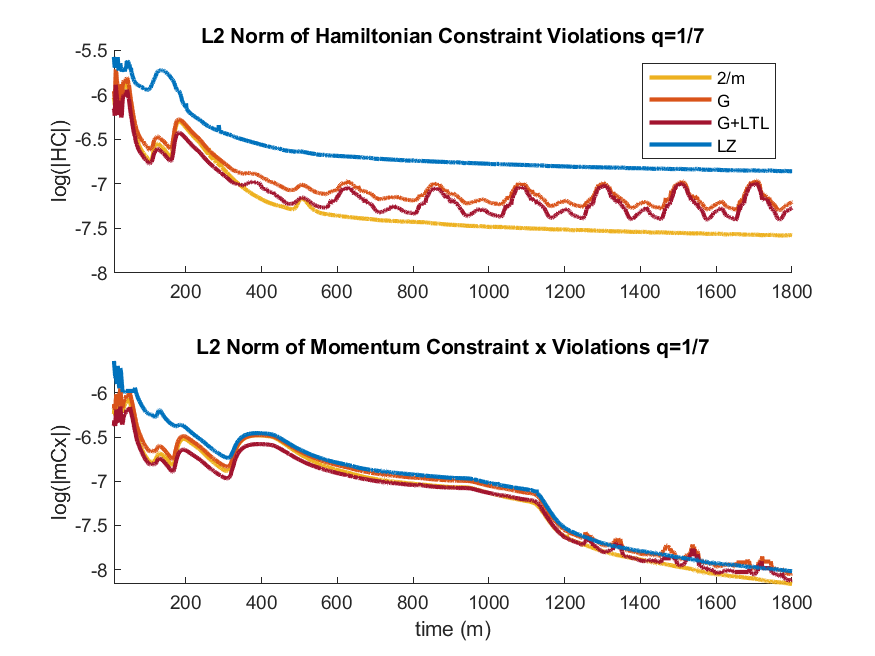}
\caption{Violations to the Hamiltonian constraint (top) and the $x-$component of the momentum constraint (bottom) for the $q=1/7$ binary using resolution n100 with the different gauge choices: G, G+LTL, $\eta=2/m$ and LZ in red, purple, yellow, and blue respectively. Both G and G+LTL settle to a similar gauge. All gauges produce acceptably-valued constraint violations.}
\label{fig:HCMC_7}
\end{figure}

The 
Hamiltonian and momentum constraint equations are integrated over a masked volume $\mathcal{V}$ and their norms are 
given by
\bea
||\mathcal{H}|| &= \sqrt{\int_\mathcal{V} \mathcal{H}^2d^3x },\\
||\mathcal{M}^i|| &= \sqrt{\int_\mathcal{V}( \mathcal{M}^i)^2d^3x }
\eea
and should be conserved~\cite{Etienne:2014tia}. Violations to this conservation are commonly used
as way to assess convergence with respect to numerical resolution. 
However, these simulations are all run in different gauges which makes it difficult to draw definitive conclusions based solely on violations to the constraints, since
different gauges can change the scales of these violations.
The constraint violations can still give us an idea of performance as well as allow us to compare different resolutions of the same gauge to ensure convergence. 
In figure \ref{fig:HCMC_7} the violations to the Hamiltonian constraint (top panel) as well as the $x-$component of the momentum constraint (bottom panel)
are shown for the $q=1/7$ simulations with $\eta=2/m$ in yellow, G in red, G+LTL in purple, and LZ in blue. 

The set up for the simulation in this subsection with $\eta=2/m$ resembles that which was used to build up the RIT Catalog~\cite{Healy:2017psd,Healy:2019jyf,Healy:2020vre}, with CFL=1/3 to achieve production speed. However, here we use eighth order finite differencing stencils instead of sixth order to simulate smaller mass-ratios, as opposed to the RIT Catalog simulations, which are mainly comparable mass.

It is pertinent to reiterate that the comparison of constraint violations cannot be used as an accurate measure of 
simulation performance between simulations with different values for $\eta$. These simulations exist in different gauges
and therefore may rescale the constraint violations, making comparisons between them an inaccurate method of ranking gauge performance.
However, one can compare the simulations that differ only in initial values (i.e. G vs. G+LTL), as well as do a ``sanity check"
to ensure that no gauge causes a dramatic increase in constraint violations. 
The pair of simulations G and G+LTL settle to approximately the same gauge, which is expected since they differ only in initial values; it also shows that initial data modifications are helpful 
only in the early part of the simulation, but do in fact reduce violations to the constraints slightly.
The constant choice, $\eta=2/m$ (in yellow) performs well throughout the course of the simulation.
The LZ gauge has Hamiltonian constraint violations that are about one order of magnitude larger than the simulations with $\eta=2/m$, G, or G+LTL, however, this may be due to a rescaling of the constraints. In general, all four simulations produce constraint violations within an acceptable range and, therefore, are considered viable candidates for production-level runs.


\begin{figure}[ht]
\includegraphics[angle=0,width=0.99\columnwidth]{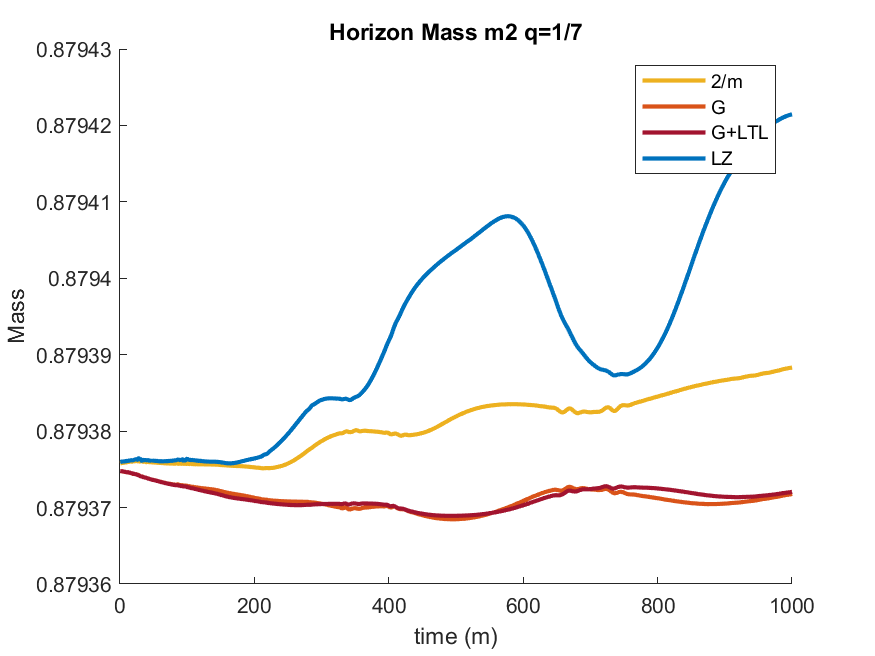}\\
\includegraphics[angle=0,width=0.99\columnwidth]{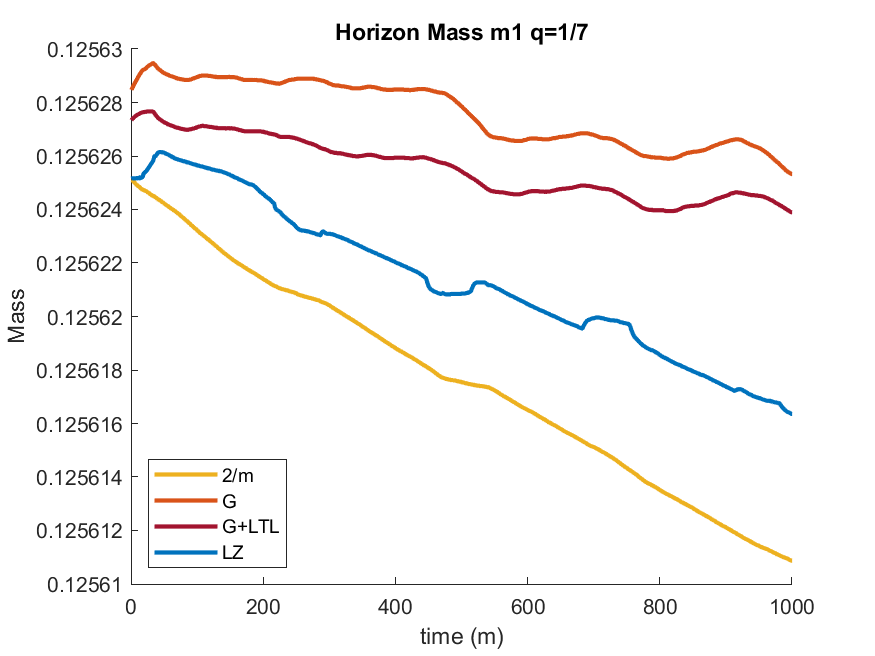}
\caption{
Horizon masses for $m_1$ and $m_2$ versus time for the $q=1/7$ n100 simulations with different choices for $m\eta$ = G, LZ, and 2/m, 
and the initial lapse, LTL. The $\eta=2/m$ and LZ gauges produce continuous growth (mass-loss) over the course of the simulation in $m_2$ ($m_1$). The G and G+LTL choices maintain both horizon masses well.}
\label{fig:mhor_7}
\end{figure}

In order to quantitatively compare the different choices for the gauge damping parameter $\eta$, it is pertinent to asses the different
gauges' effects on 
physical quantities such as horizon mass, spin, and gravitational waveforms. Fig.~\ref{fig:mhor_7} shows the mass of each black hole $m_1$ and $m_2$ measured at the horizon using LZ, G, G+LTL and $\eta=2/m$ gauges. The figures are generated using a 200-point running average to smooth fluctuations in the data and to better present a general trend without the distraction of numerical noise. In the masses, we are looking for constancy over the course of the inspiral from $t=(0-1000)$.

The top panel of Fig.~\ref{fig:mhor_7} shows the horizon mass of the large black hole $m_2$ versus simulation time. 
The G and G+LTL gauges both are able to maintain the horizon measure of mass well until the simulation approaches merger, when we see some growth in the masses. 
The constant gauge $\eta=2/m$ shows continuous growth over the course of the simulation; although the scale on which this growth occurs is small, $O(10^{-5})$, this may be indicative of an issue as the mass ratio decreases. The mass $m_2$ in the LZ gauge has relatively large oscillations over the duration of the inspiral as well as growth that is on the order of that seen in the $\eta=2/m$ simulation.

The second panel of Fig.~$\ref{fig:mhor_7}$ shows the mass of the smaller black hole, $m_1$, measured at the horizon. 
The G and G+LTL gauges both are able to maintain the horizon measure of mass well over the course of the simulation. In fact, they seem to behave the same way with G+LTL being shifted by a constant factor, indicating that the G and G+LTL settle to similar gauges since they differ only in initial data. 
The $\eta=2/m$ and LZ runs show continuous declines in their respective horizon masses of $m_1$ over the inspiral period. As we will see in the next section, this may be the symptom of a resolution issue, especially at smaller mass ratios.
Although not as vital as in smaller mass ratio systems, 
such as the $q=1/15$ or 
$q=1/32$ binaries studied in Sections 
\ref{sec:15results} and \ref{sec:32results}, here 
constancy in $m_1$ is strongly desired.

\begin{figure}[h]
\includegraphics[angle=0,width=0.99\columnwidth]{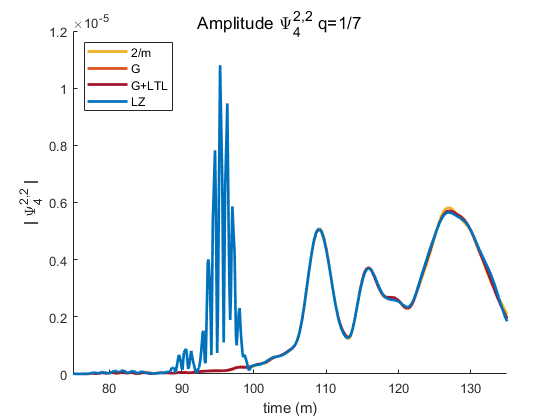}
\includegraphics[angle=0,width=0.99\columnwidth]{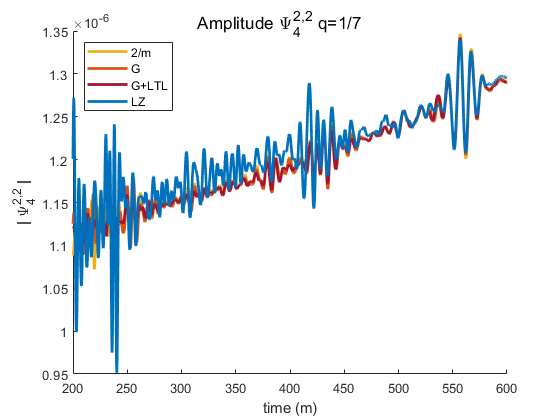}
\caption{The amplitude of the dominant (2,2)-mode of Weyl scalar $\Psi_4$ for the $q=1/7$ binary as seen by an observer situated
  $r=113m$ from the origin of coordinates.  Higher frequency noise of the LZ gauge
is apparent at $t\sim(75-135)m$ (in the top panel) as is its bounce at the next refinement level
at $t\sim(200-600)m$ (in the bottom panel). The other gauge choices G, G+LTL, and $\eta=2/m$ show no high frequency noise early on, as well as substantially reduce high frequency oscillations during inspiral (shown in the bottom panel).
}
\label{fig:Psi4_7}
\end{figure}

It is of interest to investigate the gauges' effects on the dominant $(2,2)$-mode of the
Newman Penrose Weyl scalar $\Psi_4$ since this scalar is what we use to calculate outgoing gravitational radiation. 
The top panel of figure \ref{fig:Psi4_7} shows the early part of the 
amplitude of 
$\Psi_4^{2,2}$ with respect to time. The observer sits at 
$113m$ from the origin of coordinates.
In the early part, there is a clear initial burst of noise present in 
the simulation that uses the LZ gauge which is damped by the use of G, G+LTL or $\eta=2/m$.
The second panel shows the amplitude of $\Psi_4^{2,2}$ over the timescale 
$t=(200-800)m$ with reflections at refinement boundaries visible between $(400-450)m$ and $(550-600)m$. 
This result holds and was verified for modes $(l,m)=(2,0),(2,1),(3,0),(3,1),(3,2),(3,3)$
as well.

As a final test of the efficiency of each of the gauge choices, one can consider their individual effects on the quasi-local computation of remnant recoil velocity, which is measured on the horizon of the remnant (in column 4 of Table~\ref{tab:kickq7}). This is compared to the total amount of linear momentum radiated away via gravitational waves (column 5). Using the modified gauge G improves the horizon measure of recoil velocity over the $\eta=2/m$ and LZ gauges by about 17\% and 15\% respectively. This is likely due to the fact that the G gauge damps to 1 far from the remnant, and we have already shown in \cite{Healy:2020iuc} that this leads to a more accurate measure of quasi-local linear momentum. 

\begin{table}[t]
\begin{tabular}{|lllll|}\hline
     $m\eta$&CFL&Resolution&Horizon& Radiated  \\\hline\hline
     LZ&1/3&n100&111.2450&92.6987\\
     2&1/3&n100&80.8354&94.3036 \\
     G&1/3&n100&95.1882&94.1923\\
     G+LTL&1/3&n100&95.8015&94.2094\\\hline
\end{tabular}
\caption{For the $q=1/7$ binary, the total linear momentum (km/s) calculated in two ways: (1) measured quasi-locally on the horizon averaged over $t=(2600-2800)m$, in column 4, and (2) measured by the amount radiated away in gravitational waves, in column 5. All simulations have resolution $n100$ as well as CFL=1/3 and 8th order finite differencing stencils. The gauges G and G+LTL allow for the most accurate measure of horizon recoil when compared against the radiated value.}
\label{tab:kickq7}
\end{table}

\subsection{Results for a $q=1/15$ nonspinning binary}\label{sec:15results}

Since the purpose of the G gauge is to improve simulations 
with mass ratios $q<1/10$, next we investigate a nonspinning system with mass
ratio $q=m_1/m_2=1/15$ starting at an initial
coordinate separation $D=8.5m$. The simulation was run through merger so that we can also investigate the gauges' effects
on the remnant's recoil velocity as well as the ringdown
phase of merger.

All simulations have CFL=1/3, except
the lowest resolution (n084) simulation with constant $\eta=2/m$ which
dies at about 100M with this configuration. 
The simulation runs successfully with CFL=1/4, but this is a significant change
in resolution and therefore affects all physical parameters associated
with the binary. We cannot directly compare simulations with different choices
for the CFL condition meaningfully, but have included this information
to show that the n084 $q=1/15$ simulation with constant $\eta=2/m$
works, but only with increased resolution in time. All $q=1/15$ simulations use eighth order
finite-differencing stencils in space.

As with the $q=1/7$ binary, our results for the smaller mass ratio systems will be focused on analysis of the physical quantities of the system 
such as masses, spins, and gravitational waveforms. These quantities are invariant
with respect to the gauge, and therefore can be used to measure differences 
in results using each gauge. 
Once the remnant is settled, we can compare the kicks calculated at the horizon by the isolated horizon formulae~\cite{Dreyer:2002mx} to those extracted from the radiated linear momentum at infinity. 

\begin{figure}[h]
\includegraphics[angle=0,width=0.99\columnwidth]{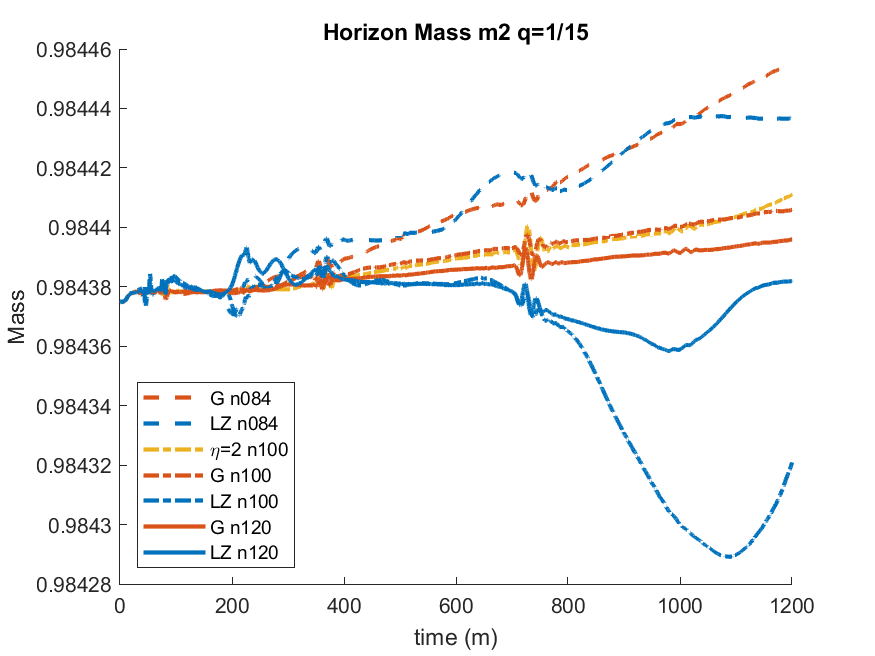}\\
\includegraphics[angle=0,width=0.99\columnwidth]{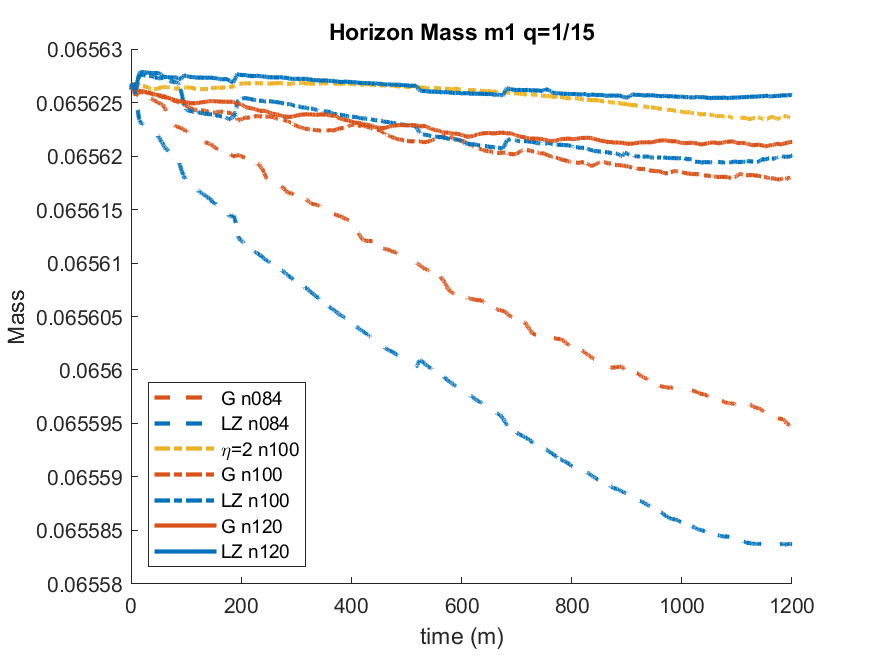}
\caption{
Horizon masses for $m_2$ (top) and $m_1$ (bottom) versus time for the $q=1/15$ simulation with the different choices for $m\eta$, G, LZ, and $\eta=2/m$. The dashed, dot-dashed, and solid lines are the low n084, medium n100, and high n120 resolutions respectively. The horizon masses $m_1$ and $m_2$ deviate from constant most dramatically when the LZ gauge is used. Both the $\eta=2/m$ and G gauges maintain the masses of the horizons well over the course of the simulation, except at the lowest resolution n084.}
\label{fig:mhor_15}
\end{figure}

The top panel of Fig.~\ref{fig:mhor_15}
shows the evolution of the horizon mass of the large black hole $m_2$ for seven simulations: G and LZ with resolutions n084, n100, and n120, and $\eta=2/m$ for n100. For the remainder of this work, the low resolution n084 will be a dashed line, n100 will be dot-dashed, and n120 will be solid. 

In Fig.~\ref{fig:mhor_15}, we can see that all gauges maintain the mass to at least O(1e-4), but the $\eta=2/m$ n100 and G n100 and n120 simulations show the best constancy in mass. The figures are generated using a 20-point running average to smooth fluctuations in the data and better present a general trend. Table~\ref{tab:slopesq15} can be used in conjunction with Fig.~\ref{fig:mhor_15}. It shows the slopes of a linear fit to the data for each simulation over the inspiral period $t=(100-1000)m$. The results of this fit are given in columns 3 and 5 of Table~\ref{tab:slopesq15} with the root mean square error 
\begin{equation}\label{eq:RMSE15}
\epsilon_i= err= \sqrt{\frac{\sum_{k=1}^N(m^k_i-\hat m^k_i)^2}{N}}
\end{equation}
where $m^k_i$ are the actual mass values for black holes labeled $i=1,2$, measured at time points $t$ and $\hat m^k_i = A t + b$.

Initially, the black holes grow due to an influx of radiation from the initial data, but then are
expected to settle to a value that remains almost constant until merger.
The top panel shows the horizon mass of the larger black hole; in the simulation using
the LZ
gauge there are low frequency oscillations later on in the inspiral, in the n084 simulation, which is reflected in the increase in $\epsilon_1$ between the linear fits of G and LZ ($\epsilon_1=0.0215\cdot10^{-6}$ vs $\epsilon_1=0.0397\cdot10^{-6}$).

 The bottom panel shows the horizon mass of the smaller black hole; in the n084 LZ gauge, 
the mass, after the settling of the initial data, declines fairly steadily, with $A=-0.3482$ and $\epsilon_2=0.7943\cdot10^{-7}$. 
This occurs in the G gauge as well, although its slope is $A=-0.2810$ with error $\epsilon_2=0.5177\cdot10^{-7}$. Although neither gauge at this resolution maintains $m_1$ very well, $m_1$ in the G gauge has a much shallower decline than the $m_1$ in the LZ gauge, as well as smaller error overall.

In the n100 simulations, $\eta=2/m$ outperforms G in terms of linearity by $<5\%$ in $m_2$ and by $>200\%$ in $m_1$, although in each case $\eta=2/m$ has higher error than G. In LZ n100 and n120, there occurs a substantial dip in the mass of $m_2$ between $t=(800-1200)m$ that is reflected in an increase in the error of the linear fits.


\begin{table}[h]
\begin{tabular}{|ll|ll|ll|}\hline
\toprule
     Gauge & Res. &\multicolumn{2}{|c|}{$m_1$}& \multicolumn{2}{|c|}{$m_2$}\\
     &&$\mathcal{A}\cdot 10^{-7}$& err.$\cdot 10^{-6}$&$\mathcal{A}\cdot 10^{-7}$ &err.$\cdot 10^{-4}$  \\\hline\hline
     G&n084&-0.2810&0.5177&0.6613&0.0215\\ 
     LZ&n084&-0.3482&0.7943&0.6462&0.0397\\ \hline
     2/m&n100&-0.0258&0.3551&0.2564&0.0131\\ 
     G&n100&-0.0683&0.2546&0.2655&0.0079\\ 
     LZ&n100&-0.0641&0.5688&-0.6248&0.1543\\ \hline
     G&n120&-0.0463&0.3206&0.1620&0.0082\\ 
     LZ&n120&-0.0245&0.2238&-0.2601&0.0467\\\hline
     \bottomrule
\end{tabular}
\caption{The slopes ($\mathcal{A}$, columns 3 and 5) of linear fits to the horizon masses $m_1$ and $m_2$ computed over the inspiral $t=100m$ to 1000 for the $q=1/15$ binary using $\eta=$G, LZ, and $2/m$. Error is calculated via root mean square error over the interval, and is shown in columns 4 and 6. The time frame is chosen so a linear fit is a reasonable approximation of the mass curves. The G and LZ simulations have 3 resolutions: n084, n100, n120, and the $\eta=2/m$ simulation has 1 resolution: n100. The G gauge produces a reliably constant value for the horizon mass, even at low resolution, whereas the mid- and low- resolution LZ simulations show large changes in mass (reflected in the slopes A) as well as higher error values overall.}
\label{tab:slopesq15}
\end{table}
It is informative to study the growth of the apparent horizon of
each black hole in numerical (radial) coordinates for each different
gauge. An extended horizon requires a large number of grid points to evaluate
quantities over its surface, but the numerical evolution ``loses'' those
points in the interior of the black hole. 
These are points that could be used to otherwise 
resolve the dynamics of the system.

\begin{figure}[h!]
\includegraphics[angle=0,width=0.99\columnwidth]{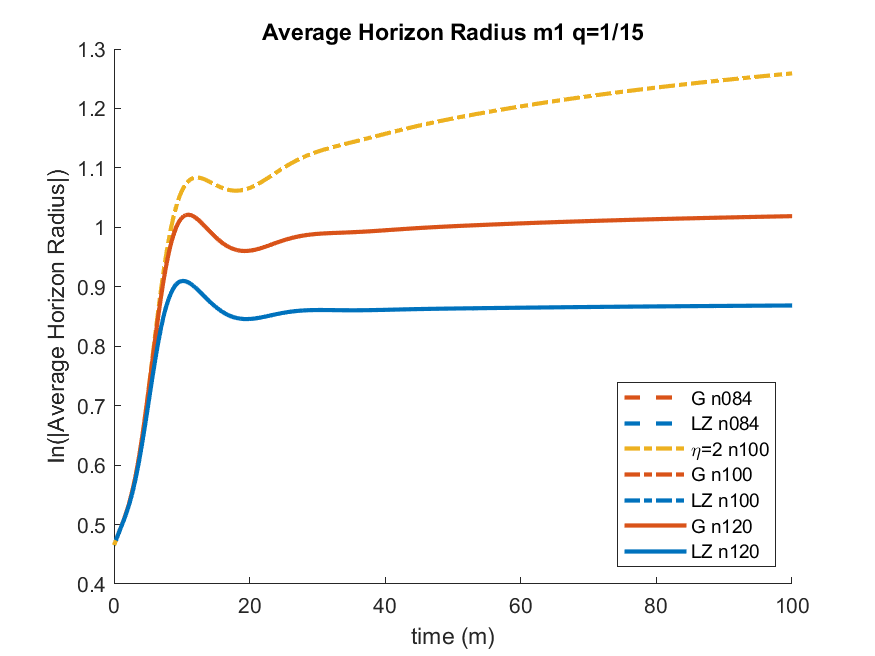}\\
\includegraphics[angle=0,width=0.99\columnwidth]{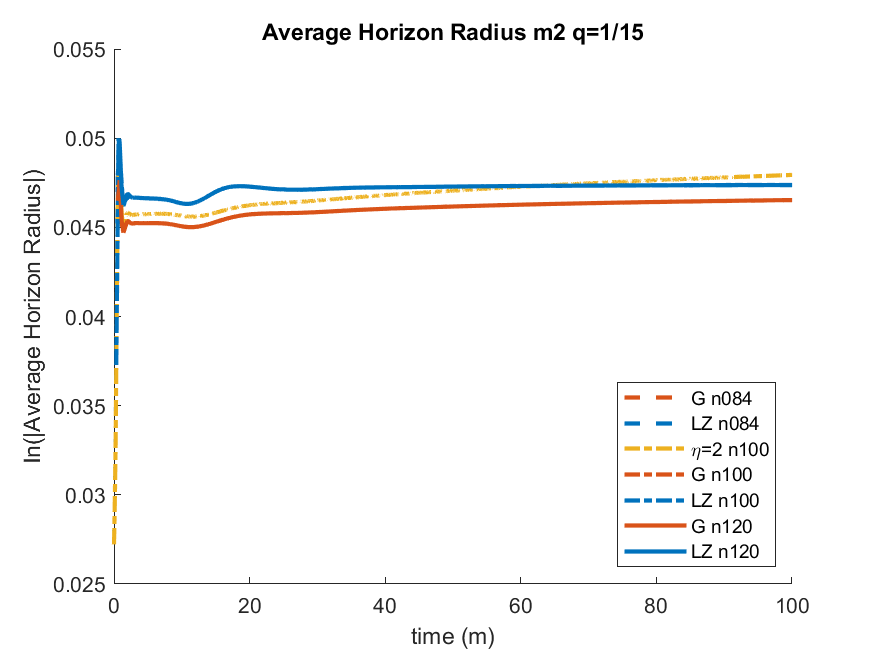}
\caption{Radial average size of each of the horizons of the $q=1/15$ binary versus evolution
time for the different gauge choices. The three resolutions using the LZ gauge lie on top of each other, as do the three resolutions using the G gauge. In both $m_1$ and $m_2$ these gauges maintain the coordinate size of the horizons well over the full inspiral. The $\eta=2/m$ is only shown in one resolution (n100), and the associated horizons $m_1$ and $m_2$ grow continuously throughout the simulation.}
\label{fig:Ravg}
\end{figure}

 The top panel of Fig.~\ref{fig:Ravg} shows the the initial growth of the larger black hole
$m_2$, which grows from $t=(0-25)m$ using all three gauge choices. This growth is due to an influx of radiation content from the initial data, and is expected. 
The growth in the small black hole (bottom panel), $m_1$, happens within the first few iterations and then immediately stabilizes. 
To minimize the loss of gridpoints in the simulation, ideally the horizon will 
grow quickly and then settle down 
to maintain its coordinate size, so this rapid stabilization is critical.

The horizon coordinate sizes for the low, medium, and high resolutions of G lie directly on top of each other in both panels of Fig.~\ref{fig:Ravg}. The same is true for all resolutions of LZ. Furthermore, the gauges G and LZ, regardless of resolution, maintain a constant coordinate size of the apparent horizons of $m_1$ and $m_2$ well over the course of the run. However, the $\eta=2/m$ simulation, shown in yellow, exhibits continuous growth of both horizons. 
This means that gridpoints are constantly being lost inside the black hole horizons, and, in the case of $m_2$, it is possible that the horizon grows so large it crosses a refinement level boundary, reducing computational accuracy and wasting resource.
While this growth does not prohibit completion of the $q=1/15$ binary at n100, further investigations should be done on its effects on more extreme mass-ratio pairs. This growth also may be related to the fact that when using $\eta=2/m$, the low resolution $q=1/15$ (n084) requires a CFL decrease from $1/3\to1/4$ in order for the simulation to be successful, whereas the other gauges G and LZ can be run at n084 with CFL of 1/3. Therefore, using a well-chosen gauge allows for an increase in computational efficiency at lower resolutions, and using a poorly chosen gauge may cause issues at low resolutions.

\begin{figure}[h!]
\includegraphics[angle=0,width=0.99\columnwidth]{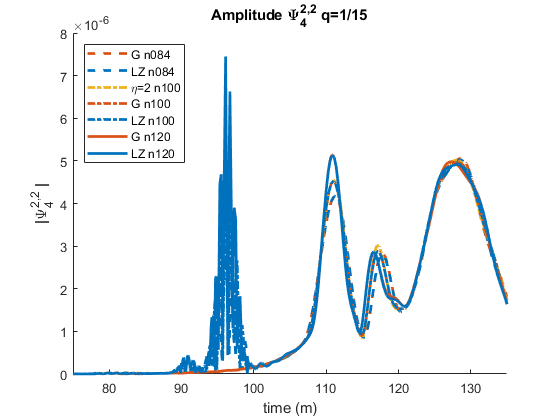}
\includegraphics[angle=0,width=0.99\columnwidth]{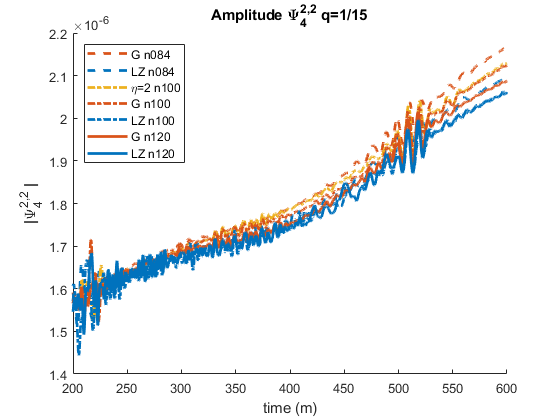}
\caption{The amplitude of the dominant mode of $\Psi_4$ for the $q=1/15$ binary extracted at an observer location of
  $r=113m$. The waveform is shown for G and LZ in three resolutions and $\eta=2/m$ in one. In the top panel, higher frequency noise of the LZ gauge early on
at $t\sim(75-135)m$ and its bounce at the next refinement level
in the second panel from $t\sim(200-600)m$ is apparent. 
}
\label{fig:Psi4}
\end{figure}

An important and physically relevant method of assessing the accuracy and effectiveness of different gauges is to look at the early behavior of gravitational waveforms as seen by an
observer far from the binary orbit. In our case, we will consider an observer sitting at $r=113m$ from the origin of coordinates.
Fig.~\ref{fig:Psi4}
displays the amplitude of the leading waveform mode $\ell=2,m=2$ of the Weyl scalar $\Psi_4$. 
In the top panel the amplitude of the waveform from $t=(75-135)m$ is shown. One can observe high frequency noise at $t\sim(90-100)m$ produced by the LZ gauge choice, in blue, that was also present in the $q=1/7$ waveform (as shown in Fig.~\ref{fig:Psi4_7}). The noise has higher amplitude in the n100 (dot-dash) LZ simulation than the n084 (dash) LZ simulation, this is not indicative of improvement with decreasing resolutions, but is instead due to the n084 resolution under-resolving the grid. This noise is eradicated solely by choosing either G or $\eta=2/m$ for the gauge, which is also consistent with what was found in Fig.~\ref{fig:Psi4_7}.

The second panel of Fig.~\ref{fig:Psi4} shows a later time in the binary's evolution
(from  $t\sim(200-600)m$). High frequency noise during this period of the inspiral, present even in the highest resolution (n120) simulation using the LZ gauge, is significantly damped when using the G or $2/m$ choice for $\eta$ instead of the typical LZ.
At $t\sim(500m-550)m$, there is an increase in the amplitude of these oscillations across all simulations. This corresponds to the bounce of noise over a grid refinement level back to the observer's location.
This effect is also damped when G or $\eta=2/m$ are chosen,
thus confirming the benefits of the introduction of the new gauges. We verified
that similar features appear in other next to leading order modes
$(\ell,m)=(2,1),(2,0),(3,0),(3,1),(3,2),(3,3)$, etc.

Table~\ref{tab:kickq15} contains the remnant quantities (energy, angular momentum, and linear momentum) of the $q=1/15$ binary including a three-point extrapolation to infinity ($n\infty$) as well as convergence order in the cases where we use three resolutions and have convergence. Columns 3-5 show the mass, and angular and linear momentum measured on the horizon of the black hole using the isolated horizon formulae. Columns 6-8 are the same quantities but calculated from the energy, angular, and linear momentum carried away by gravitational waves off to an infinite-location observer. Excellent agreement is observed between radiation and horizon measurements for the final mass and spin. In general, the horizon linear momenta are measured well at the typical production resolutions of n100 and n120 using the G gauge, coming within 12\% of the radiated measure for n100 and 28\% for n120. Compare this with the LZ gauge, which is off wildly (by 716\% using n100) or the $\eta=2/m$ gauge which is off by 35\% using n100. 

Since $\eta=1/m$ gives a more accurate measure of recoil velocity than $\eta=2/m$, and G damps to 1 far from the center of coordinates, we expected improvement between $\eta=2/m$ and G. Additionally, the strong improvement between G and LZ may prove critical when considering extremely small mass ratios, such as the $q=1/32$ or $q=1/64$ binaries considered here. 
\begin{table*}[t]
\begin{tabular}{|cc|ccc|ccc|}\hline
\toprule
     &&\multicolumn{3}{|c|}{Horizon}& \multicolumn{3}{|c|}{Radiated}\\
     Gauge&Resolution&$E/m$&$J/m^2$&$L$(km/s)&$E/m$&$J/m^2$&$L$(km/s)\\\hline
     &n084&0.9949&0.1863&178.7939&0.9950&0.1889&	31.5722\\
     &n100&0.9949&0.1870&36.3859&0.9950&0.1880&	32.9681\\
     G&n120&0.9949&0.1870&44.79436&0.9949&0.1877&35.0123\\
     &$n\infty$&0.9950&0.1870&44.3255&0.9949&0.1874&35.1600\\
     &CO&4.64&15.50&15.52&3.51&5.05&0.8630\\\hline\hline
     &n084&0.9949	&	0.1865&232.3947&	0.9950&	0.1881&31.0327\\
     &n100&0.9948	&	0.1862&241.2582&	0.9949&	0.1876&34.3328\\
     LZ&n120&0.9949&	0.1872&43.4521&	0.9949&	0.1874&35.0915\\
     &$n\infty$&0.9948	&	0.1864&-&	0.9949&	0.1874&35.3179\\
     &CO&3.11&	2.78&-&	5.88&	7.12&8.06\\\hline \hline
     $2/m$&n100&0.9949&0.1870&21.1410&0.9950&0.1884&	32.7075\\\hline
     \bottomrule
\end{tabular}
\caption{For the $q=1/15$ binary, the total mass, angular, and linear momentum (1) measured by the amount radiated away in gravitational waves, in columns 3-5, and (2) measured quasi-locally on the horizon, in columns 6-8. Extrapolations to infinite resolution ($n\infty$) and convergence order (CO) are included where applicable. All simulations have resolution specified in column 3 as well as CFL=1/3 and 8th order finite differencing stencils. A dash indicates no convergence was found.}
\label{tab:kickq15}
\end{table*}

\subsection{Results for a $q=1/32$ nonspinning binary} \label{sec:32results}

We also performed an in-depth study on a smaller nonspinning binary with a mass-ratio of $q=1/32$ and an initial binary separation of $d=8m$.  It uses a Courant factor of $1/4$ as well as 8th order finite differencing stencils in space, and was run using three resolutions (n084, n100, n120), with both the G and LZ gauges as well as one resolution (n100) of $\eta=2/m$. 

\begin{figure}[h!]
\includegraphics[angle=0,width=0.99\columnwidth]{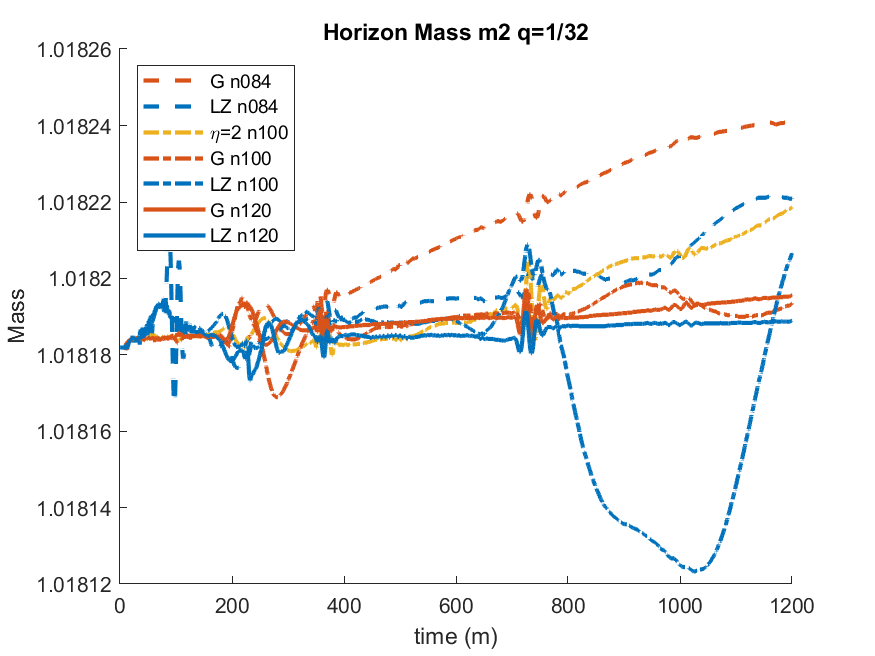}\\
\includegraphics[angle=0,width=0.99\columnwidth]{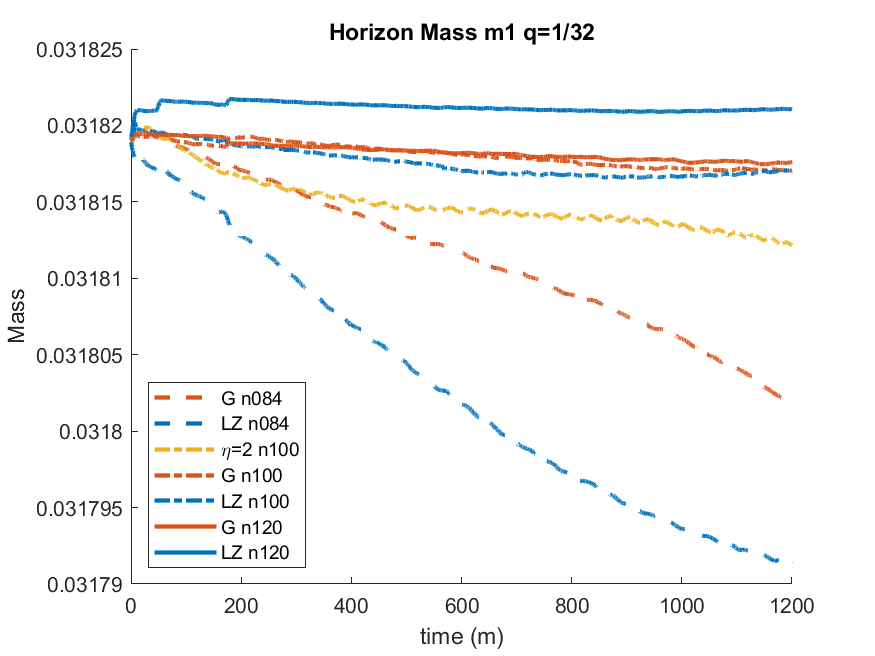}
\caption{
Horizon masses for $m_2$ (top) and $m_1$ (bottom) versus time for the $q=1/32$ simulation with the different choices for $m\eta$, G, LZ, and $\eta=2/m$. The dashed, dot-dashed, and solid lines are the low n084, medium n100, and high n120 resolutions respectively. The horizon masses $m_1$ and $m_2$ deviate from constant most dramatically when the LZ gauge is used. Both the $\eta=2/m$ and G gauges maintain the mass of the horizon well over the course of the simulation, except at the lowest resolution n084. }
\label{fig:mhor_32}
\end{figure}

Fig.~\ref{fig:mhor_32} shows the horizon masses for the seven $q=1/32$ simulations during inspiral from $t=0m$ through $t=1200m$, which is just before merger. 
In these figures, we are again looking for the masses to be held constant (after settling down initially) for the duration of the inspiral, as this will indicate a more accurate computation of horizon masses. The lowest resolution (n084) simulations using the G and LZ gauges show the most growth in $m_2$ (top panel) or mass-loss in $m_1$ (bottom panel), while the G n120 simulation holds both masses most constant over time. The n100 LZ gauge shows a relatively large dip in $m_2$ mass between $t=(800-1200)m$, which is consistent with our findings for the $q=1/15$ binary. The $\eta=2/m$ n100 simulation holds $m_2$ constant until $t=800m$ and then begins to grow at the same rate as the LZ n084 simulation. In $m_1$, $\eta=2/m$ begins with mass loss from $t=(0-200)m$ then stays relatively constant until $t=1100m$ where it begins to lose mass again. While not prohibitive to the completion of this particular simulation, this mass loss in the small black hole might pose issues at lower resolutions or mass ratios. 

To assess continuity of the horizon mass parameters quantitatively, we performed a linear fit to the data for each simulation over the inspiral period $t=(100-1000)m$. The results of this fit are given in columns 3 and 5 of Table~\ref{tab:slopes} with the root mean square error as in~(\ref{eq:RMSE15}).

The LZ n084 $m_2$ and $m_1$ have slopes of $A_1=0.2332\cdot10^{-4}$ and $A_2=-0.2529\cdot 10^{-7}$ (respectively), whereas G has slopes $A_1=0.6370\cdot 10^{-4}$ and $A_2=-0.1354\cdot 10^{-7}$. The slope of $m_2$ is most constant using LZ, as well as has the lower error. However, the more difficult to resolve black hole is the smaller horizon, whose mass is better computed using the G gauge. 

As resolution increases, so does the stability of the masses of each black hole; using G at n100 provides horizon mass measurements that are closer to constant as well as using $\eta=2/m$, however, because of the dip using LZ at $t=(800-1200)m$, the slopes for LZ n100 are similar to those of n084. The G n100 gauge matches the G n120 gauge well throughout most of the inspiral period, whereas $\eta=2/m$ has growth in the mass that is superlinear and is reflected in the increase in root mean squared error between G (at $\epsilon_1=0.0079\cdot10^{-4}$ and $\epsilon_2=0.2546\cdot 10^{-7}$) and $\eta=2/m$ (at $\epsilon_1=0.0131\cdot10^4$ and $\epsilon_2=0.3551\cdot 10^{-7}$). A full table of the mean slopes of the horizon masses can be found in table~\ref{tab:slopes}.

\begin{table}[h!]
\begin{tabular}{|ll|ll|ll|}\hline
\toprule
     Gauge & Res. & \multicolumn{2}{|c|}{$m_1$}&\multicolumn{2}{|c|}{$m_2$}\\
     &&$\mathcal{A}\cdot 10^{-7}$& err.$\cdot 10^{-6}$&$\mathcal{A}\cdot 10^{-7}$ &err.$\cdot 10^{-4}$  \\\hline\hline
     G&n084&-0.1354&0.0977&0.6370&0.0222\\ 
     LZ&n084&-0.2529&0.5638&0.2332&0.0275\\ \hline
     2/m&n100&-0.0410&0.4051&0.2620&0.0407\\ 
     G&n100&-0.0231&0.1507&0.1739&0.0402\\ 
     LZ&n100&-0.0310&0.2049&-0.5344&0.1673\\ \hline
     G&n120&-0.0163&0.1054&0.0659&0.0171\\ \hline
     \bottomrule
\end{tabular}
\caption{The slopes ($\mathcal{A}$, columns 3 and 5) of linear fits to the horizon masses $m_1$ and $m_2$ computed over the inspiral $t=(0-1000)m$ for the $q=1/32$ binary using $\eta=$G, LZ, and $2/m$. Error is calculated via root mean square error over the interval, and is shown in columns 4 and 6. The time frame is chosen so a linear fit is a reasonable approximation of the mass curves. The G and LZ simulations have 3 resolutions: n084, n100, n120, and the $\eta=2/m$ simulation has 1 resolution: n100. The G gauge provides an effective gain in resolution of a factor of 2 in the horizon mass.}
\label{tab:slopes}
\end{table}

\begin{figure}[h!]
\includegraphics[angle=0,width=0.99\columnwidth]{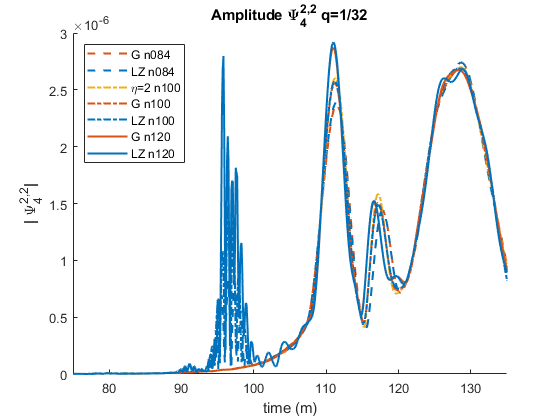}
\includegraphics[angle=0,width=0.99\columnwidth]{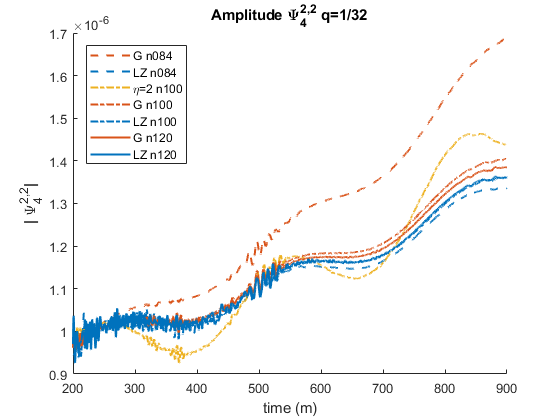}
\caption{The amplitude of the $\Psi_4$ waveform of the $q=1/32$ binary extracted at an observer located at
  $r=113m$. The top panel shows the early part of the waveform from $t=(75-135)m$, and the bottom panel shows the inspiral period from $t=(200-900)m$. Higher frequency noise of the LZ gauge
is apparent at $t\sim(90-100)m$, which is damped by the G and $\eta=2/m$ gauges. Refinement boundary reflections of the high frequency noise are apparent in the second panel, which shows a longer inspiral.
 }
\label{fig:Psi4q32}
\end{figure}

In Fig.~\ref{fig:Psi4q32}, the amplitude of the gravitational wave scalar $\Psi_4^{2,2}$ is plotted versus time for all seven simulations. The top panel shows the early part of the waveform, from $t=75-135m$. Between $t=90-100m$, there is noise in the LZ simulations that is damped by the G and $\eta=2/m$ simulations across all resolutions which is consistent with our findings for the $q=1/15$ and $q=1/7$ binaries. This means that smaller mass-ratios benefit significantly from the use of an adaptive gauge like G or a constant choice like $\eta=2/m$. However, recall that lower resolutions of $\eta=2/m$ may require at least increases in time resolution (via CFL) which are not necessarily computationally practical, especially as mass-ratios decrease.

The bottom panel of Fig.~\ref{fig:Psi4q32} shows the inspiral period of $\Psi_4^{2,2}$ from $t=(200-900)m$. Compare the low resolution G (n084, red dashed) and the n100 (blue, dot dashed) resolution of the same gauge between 400m and 450m. The n100 curve shows noise reflected at the mesh refinement boundary, but the G n084 curve seems to damp this noise. This is due to n084 under-resolving the grid, which is also the cause of the increase in initial noise between LZ n084 and n100. The high-frequency oscillations occurring from the subsequent reflection between $t=(500-600)m$ also do not appear in the G n084 simulation, whereas they do in the G n100 and G n120 simulations, which supports the hypothesis that n084 has too few gridpoints to properly resolve the system. 

The LZ gauge in n084 (blue, dashed) shows high frequency oscillations over the course of the inspiral, which are somewhat damped by increasing the resolution to n100 (blue, dot-dashed). The G n100 simulation (red, dot-dashed) exhibits less oscillatory behavior when compared to the waveform in the other gauges, $\eta=2/m$ and LZ, at resolution n100. 

\begin{figure}[h!]
\includegraphics[angle=0,width=0.99\columnwidth]{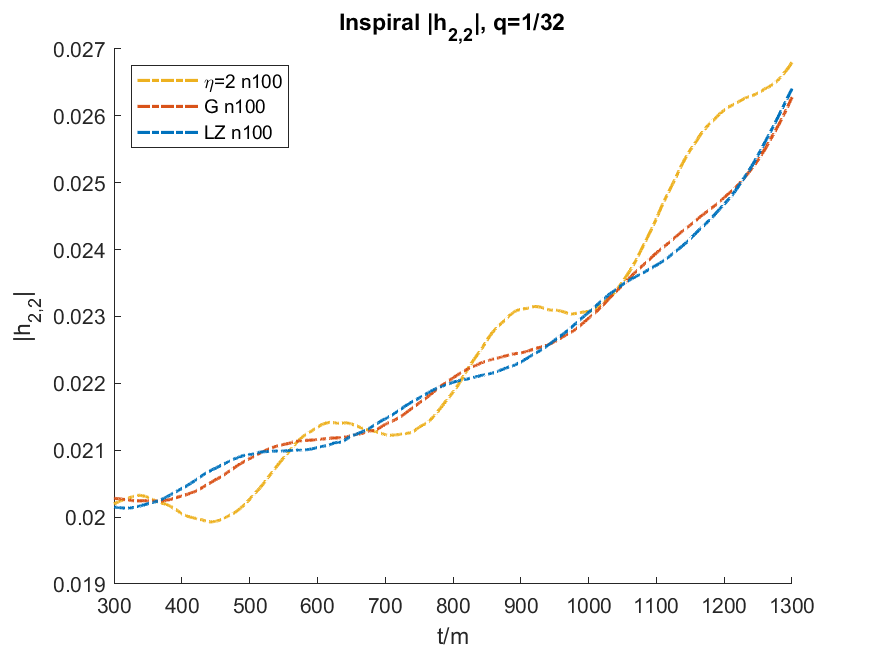}\\
\includegraphics[angle=0,width=0.99\columnwidth]{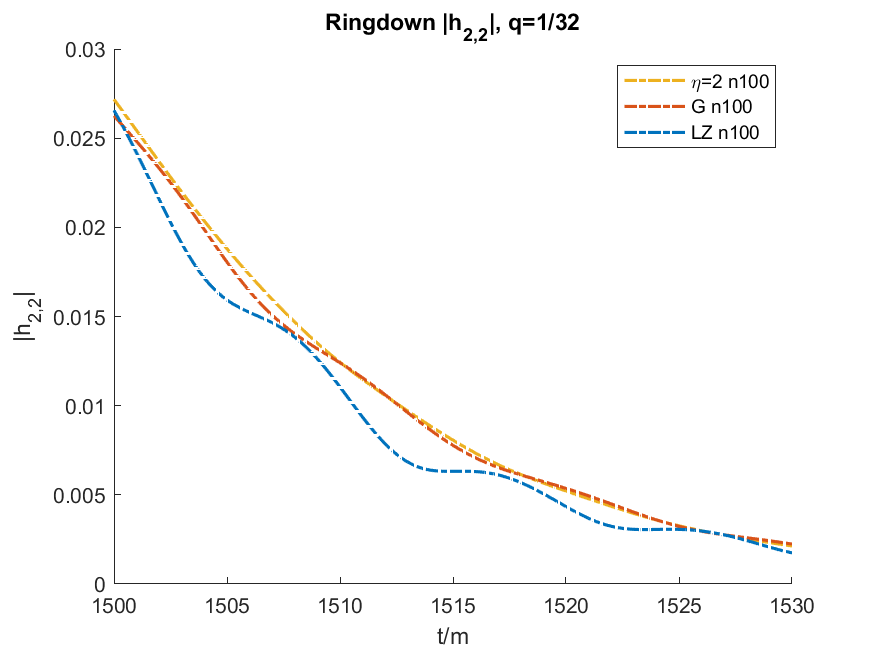}
\caption{Inspiral (top) and ringdown (bottom) periods of the gravitational wave strain $h_{2,2}$ of the mid-resolution simulations of the $q=1/32$ binary using the G, LZ, and $\eta=2/m$ gauges. In the inspiral, the waveform using the $\eta=2/m$ gauge exhibits oscillatory behavior which is not present in either of the variable gauges. However, post-merger, the waveform using LZ exhibits oscillations, whereas the G and $\eta=2/m$ waveforms do not. This suggests that a variable gauge is useful during inspiral, and a more constant gauge is most effective post-merger. }
\label{fig:Mergeq32}
\end{figure}

For smaller mass ratios, comparing the waveforms in the near-merger inspiral and ringdown phases is also a good method of quantifying gauge performance.
Fig.~\ref{fig:Mergeq32} shows two different sections of the amplitude of the gravitational wave strain $h_{2,2}$ for the $q=1/32$ binary, using the G, LZ and $\eta=2/m$ gauges at resolution n100, extrapolated to an observer at $\infty$. The merger times for all simulations are matched for easy comparison. The top panel of Fig.~\ref{fig:Mergeq32} shows the inspiral portion of the waveform, from $t=(300-1300)m$ and the bottom panel shows the ringdown period post-merger from $t=(1500-1530)m$.

In Fig.~\ref{fig:Mergeq32}, we are interested in comparing the effects of the different gauges on low frequency oscillations before and after merger. Using the constant choice $\eta=2/m$ introduces low frequency oscillations in the inspiral portion (top panel) of the strain $h_{2,2}$ that are damped by choosing an alternative, variable gauge such as G or LZ. This suggests that using a gauge with peaks at the black holes resolves the binary's dynamics well in the strong field region, effectively damping oscillations in the gauge that propagate from the black holes. 
The ringdown phase (bottom panel) shows significant oscillations in gravitational waveforms that use the LZ gauge. Using $\eta=2/m$ or G produces waveforms without these oscillations, which suggest that the low frequency oscillations come from the choice of gauge itself. We are currently investigating a gauge that uses G for the inspiral and then flattens to a near-constant $\eta=1/m$ post-merger, i.e. Eq.~(\ref{eq:etaG}) with $n=2$.

As a follow up to \cite{Healy:2020iuc}, we would like to investigate the effects of different gauge choices on the horizon quantities of small mass ratio binaries as well. Table~\ref{tab:kickq32} shows the results of this study. It includes all 7 runs with mass ratio $q=1/32$ (3 resolutions for G and LZ, and one for $\eta=2/m$) as well as a 3-point extrapolation to infinity in the row labeled $n\infty$ and the order of convergence (CO) where they exist. The Horizon columns show the results of the energy and angular momentum calculated using the isolated horizon formulae, and the Radiated columns show the results of calculating the same three quantities as carried away by gravitational waves. 

Our new gauge G does a good job of measuring mass and angular momentum on the horizon when compared to its radiated counterpart. The horizon energy and radiated energy converge to the same value (within $10^{-4}$), as do the horizon and radiated angular momentum.

However, at the resolutions we used for these simulations, G, and our typical variable gauge, LZ, do a very poor job of measuring kick on the horizon when compared to the linear momentum carried away by gravitational waves. For this reason, we have chosen to omit these results from Table~\ref{tab:kickq32}. It is worth mentioning that G performs slightly better, at least reaching the correct order by the highest resolution  - 20.85 km/s measured on the horizon vs. 9.94 km/s measured at infinity, where LZ estimates 219.96 km/s instead of 9.73 km/s.
For both G and LZ, the approximation does improve with resolution, however for an accurate measurement of horizon linear momentum we need to increase the resolution again to n144. 
The constant gauge $\eta=2/m$, was only simulated for one resolution, n100, of $q=1/32$; because of the increase in CFL that was required for the lowest resolution $q=1/15$, we have discounted $\eta=2/m$ as a viable option.

\begin{table*}[t]
\begin{tabular}{|cc|cc|ccc|}\hline
\toprule
     &&\multicolumn{2}{|c|}{Horizon}& \multicolumn{3}{|c|}{Radiated}\\
     Gauge&Resolution&$E/m$&$J/m^2$&$E/m$&$J/m^2$&$L$ (km/s)\\\hline
     &n084&0.9979&0.0985&0.9979&0.0982&	8.4673\\
     &n100&0.9979&0.0972&0.9979&0.0979&	9.7856\\
     G&n120&0.9979&0.0977&0.9979&0.0978&9.9423\\
     &$n\infty$&0.9978&0.0976&0.9979&0.0977&9.9634\\
     &CO&2.00&4.42&4.82&6.04&11.68\\\hline\hline
     &n084&0.9979&0.0974&0.9979&0.0978&	7.3265\\
     &n100&0.9979&0.1002&0.9979&0.0978&	9.5137 \\
     LZ&n120&0.9979&0.0967&0.9979&0.0976&9.7311\\
     &$n\infty$&0.9979&0.0973&0.9980&0.0971&9.7551\\
     &CO&4.23&15.04&0.90&0.06&12.66\\\hline \hline
     $2/m$&n100&0.9979&0.0976&0.9979&0.0978&	9.3891\\\hline
     \bottomrule
\end{tabular}
\caption{For the $q=1/32$ binaries using G, LZ, and $\eta=2/m$, the resolutions are given in column 2. The horizon quantities (energy and angular momentum) are in columns 3 and 4, and the corresponding radiated quantities are in columns 5-7. All simulations use CFL=1/4 and 8th order finite differencing stencils. The G gauge does a good job of measuring the horizon mass and spin, however, the horizon linear momentum requires high resolution to approach the radiated linear momentum.}
\label{tab:kickq32}
\end{table*}

\subsection{Results for a $q=1/64$ nonspinning binary}
\label{sec:q64128results}
We also performed a study on an extremely small mass ratio, nonspinning binaries with $q=1/64$ and $q=1/128$ and an initial binary separation of $d=7m$. The simulations use a Courant factor of $1/4$ as well as 8th order finite differencing stencils. The $q=1/64$ was run with two resolutions (n084, n100) in the LZ gauge and one resolution (n100) in the G gauge. As a follow up to \cite{Lousto:2020tnb}, the $q=1/128$ n100 simulation in the LZ gauge generated for that work is included here to compare with preliminary results for the n100 G gauge.

\begin{figure}[h]
\includegraphics[angle=0,width=0.99\columnwidth]{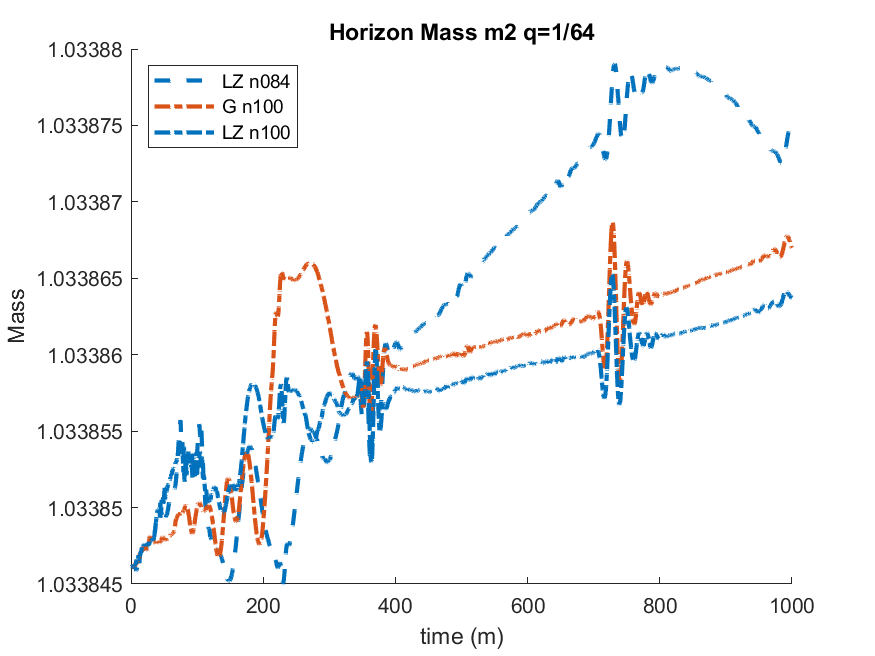}\\
\includegraphics[angle=0,width=0.99\columnwidth]{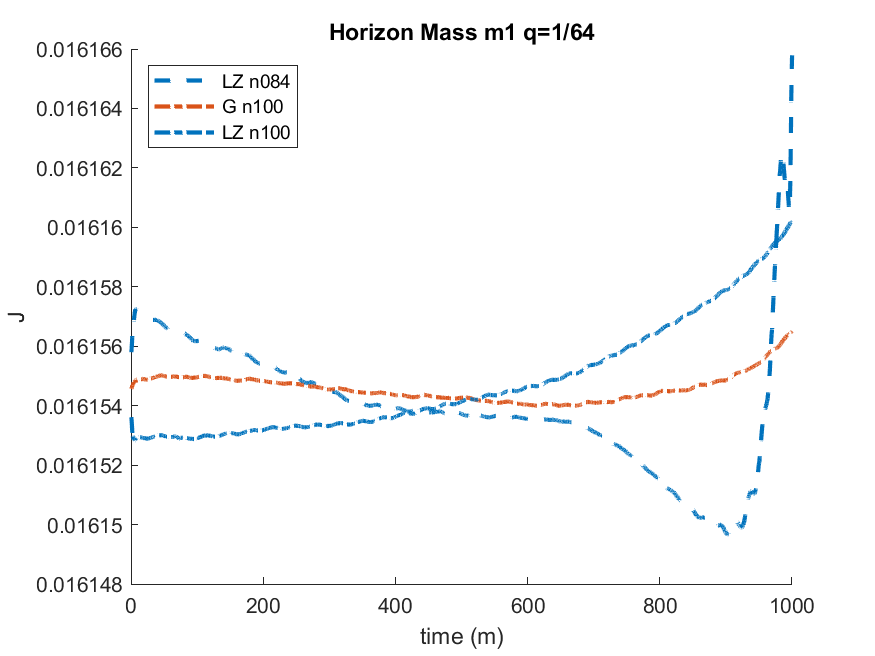}
\caption{
Horizon masses for $m_1$ and $m_2$ versus time for the $q=1/64$ simulation. The masses in the LZ gauge are shown in two resolutions n084 and n100, and the masses in the G gauge are shown in only n100. The masses using G are more constant than their counterparts that use the LZ gauge, especially in the case of the small black hole $m_1$.  }
\label{fig:mhor_64}
\end{figure}

In Fig.~\ref{fig:mhor_64}, the masses measured on the horizon of the black holes (top: $m_2$, bottom: $m_1$) are shown for $q=1/64$. The LZ gauge is in blue and the G gauge is in red with different resolutions denoted by different line types. In $m_2$, both gauges produce masses that are similarly constant at our typical production resolution of n100. The resolution n084 is quite a bit under-resolved for this extremal simulation, and therefore the LZ gauge does not do a good job of maintaining constancy in either $m_1$ or $m_2$. However, the smaller black hole $m_1$ is more difficult to resolve, and using the G gauge at n100, its mass is held more constant than using the LZ gauge at the same resolution. 

\begin{figure}[h]
\includegraphics[angle=0,width=0.99\columnwidth]{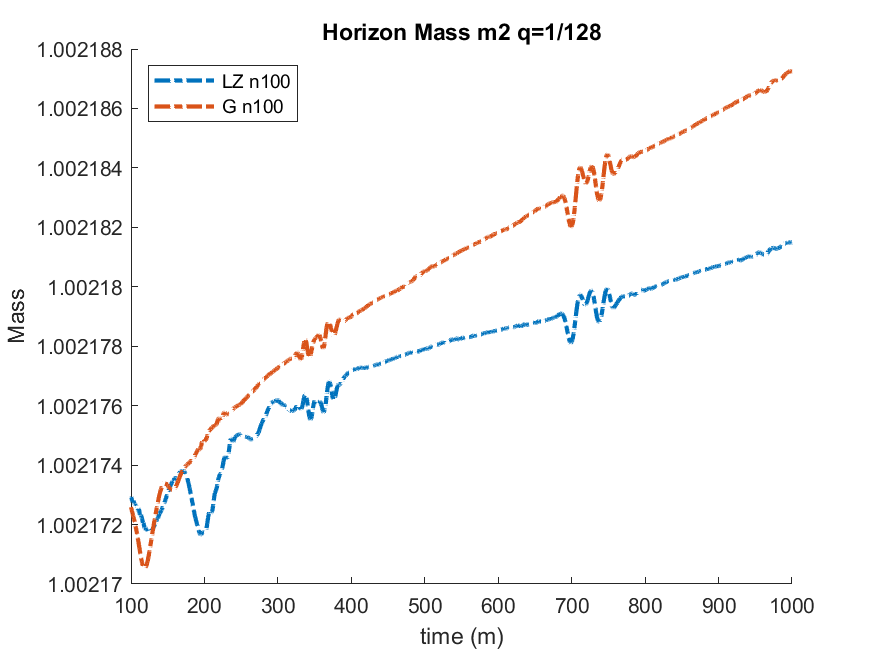}\\
\includegraphics[angle=0,width=0.99\columnwidth]{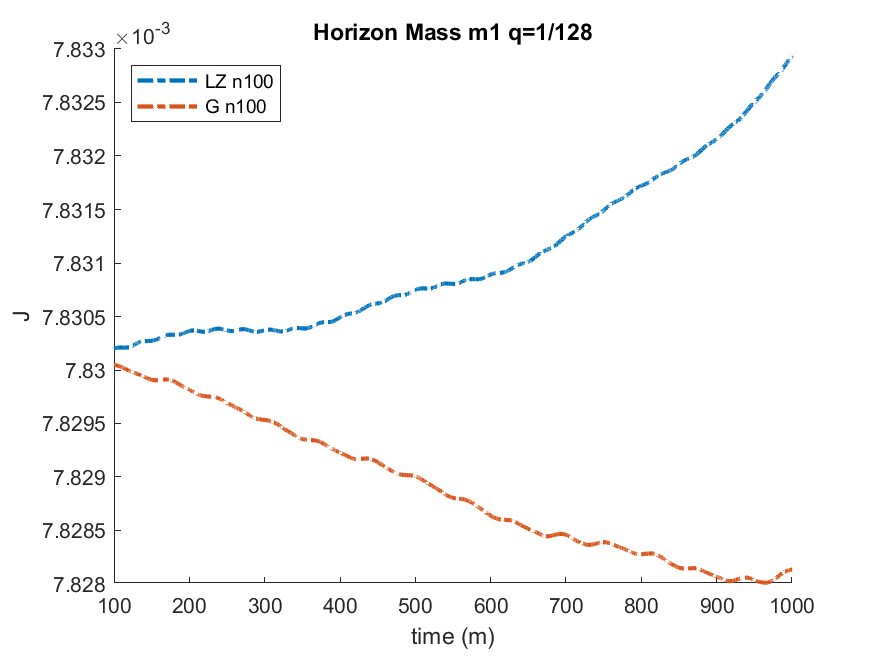}
\caption{
Horizon masses for $m_1$ and $m_2$ versus time for the $q=1/128$ simulation. The masses in both the LZ and G gauges are shown in only n100. The large black hole $m_2$ using the LZ gauge is held slightly more constant than its counterpart that uses the G gauge, however the LZ gauge simulation has significantly more noise than the G gauge one.  }
\label{fig:mhor_128}
\end{figure}

For the more extreme mass ratio binary, $q=1/128$, we will consider only the early part of the n100 simulation in the G gauge; the simulation was not run through merger. This will be compared to a full simulation of the binary in the LZ gauge since it was completed for \cite{Lousto:2020tnb}. These runs are very computationally difficult to do, but comparing the results of the early part of the simulations will give us a good idea of the benefits of using G instead of LZ here. Fig. \ref{fig:mhor_128} shows both horizon masses versus simulation time in the G and LZ gauges in one resolution (n100). Since n100 is still under-resolved for the $q=1/128$ binary, neither $m_1$ nor $m_2$ is held constant. However, $m_2$ in the LZ gauge has slightly less mass-gain than in the G gauge, whereas the G gauge has a significant reduction in noise over the LZ gauge. Neither gauge prevents mass loss/gain in $m_1$ - in order to see benefits of the G gauge it is likely we would need to increase resolution by at least a factor of 1.2 (to n120). 

\begin{figure}[h!]
\includegraphics[angle=0,width=0.99\columnwidth]{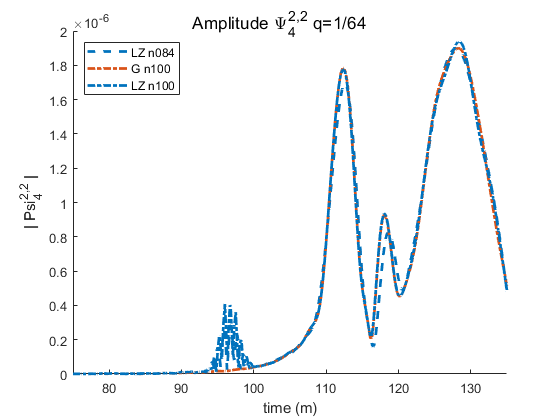}
\includegraphics[angle=0,width=0.99\columnwidth]{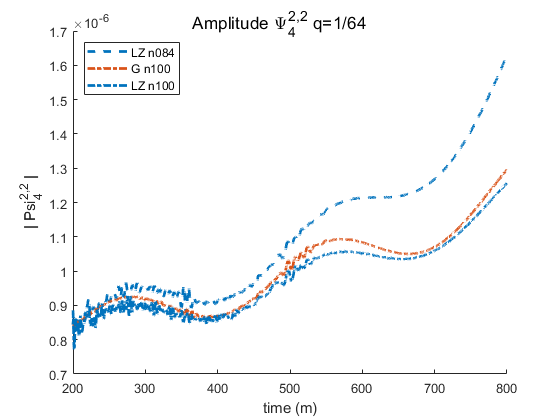}
\caption{The amplitude of the $\Psi_4$ waveform of the $q=1/64$ extracted at an observer located at
  $r=113m$. The top panel shows the early part of the waveform from $t=(75-135)m$, and the bottom panel shows the inspiral period from $t=(200-900)m$. Higher frequency noise of the LZ gauge
is apparent at $t\sim(95-100)m$, which is damped by the G and $\eta=2/m$ gauges. Refinement boundary reflections of the high frequency noise are apparent in the second panel, which shows a longer inspiral.
 }
\label{fig:Psi4q64}
\end{figure}

Fig. \ref{fig:Psi4q64} shows the amplitude of the gravitational wave scalar $\Psi_4^{2,2}$ versus simulation time for the three different runs with mass ratio $q=1/64$. The top panel shows the early part of the inspiral, and the numerical noise present when the LZ gauge is chosen is visible again between $t=(95-100)m$, however it has decreased substantially in amplitude. Using the G gauge damps this noise completely, as well as reduces other oscillations present in the LZ gauge simulation with n100 resolution (blue, dot-dashed) at $t=150m$. 

The bottom panel shows the inspiral period of $\Psi_4^{2,2}$ from $t=(200-800)m$. The simulation with the LZ gauge in resolution n100 has high frequency oscillations between $t=(200-500)m$ which are nonphysical. These are damped away almost completely by making the choice of G for $m\eta$. There are reflections of noise at the grid refinement boundaries visible between $t=(300-400)m$ and $t=(500-600)m$. 

\begin{figure}[ht]
\includegraphics[angle=0,width=0.99\columnwidth]{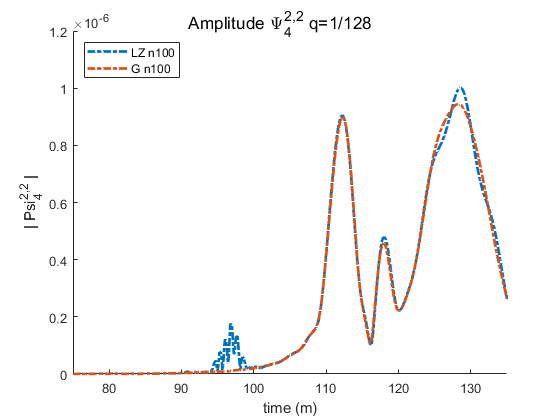}
\includegraphics[angle=0,width=0.99\columnwidth]{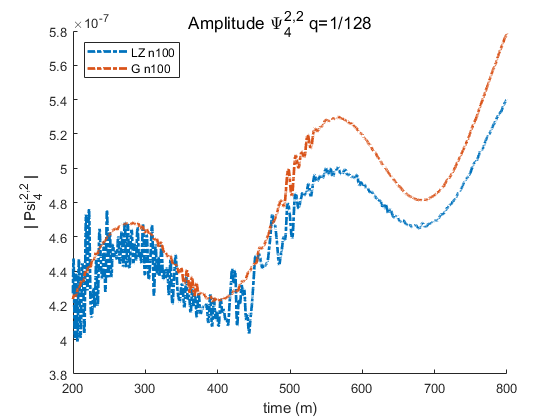}
\caption{The amplitude of the $\Psi_4$ waveform of the $q=1/128$ extracted at an observer located at
  $r=113m$. The top panel shows the early part of the waveform from $t=(75-135)m$, and the bottom panel shows the inspiral period from $t=(200-800)m$. Higher frequency noise of the LZ gauge
is apparent at $t\sim(95-100)m$, which is damped by the G and $\eta=2/m$ gauges. Refinement boundary reflections of the high frequency noise are apparent in the second panel, which shows a longer inspiral.
 }
\label{fig:Psi4q128}
\end{figure}

Similarly, we can consider the effects of the different gauges on the inspiral period of the $q=1/128$ binary. In contrast to the horizon masses, when the gravitational waveform is considered, the G gauge shows obvious improvement over LZ. Fig. \ref{fig:Psi4q128} shows the inspiral from $t=(75-135)m$ in the top panel and $t=(200-800)m$ in the bottom panel using both G and LZ in resolution n100. In the top panel, between $t=(95-100)m$, we can see an initial burst of nonphysical noise that is consistent with our results for all other mass-ratios. This is damped by choosing the G gauge, which is hugely beneficial for such a demanding simulation. 

The bottom panel of Fig.~\ref{fig:Psi4q128} shows the inspiral period of $\Psi_4^{2,2}$ up to just before merger. In the LZ gauge, there is a substantial amount of high frequency noise present in the waveform until about $t=600m$. Using the G gauge, this noise is almost completely damped, producing a cleaner and more accurate gravitational waveform for no increase in computational expense. 

\begin{figure}[h!]
\includegraphics[angle=0,width=0.99\columnwidth]{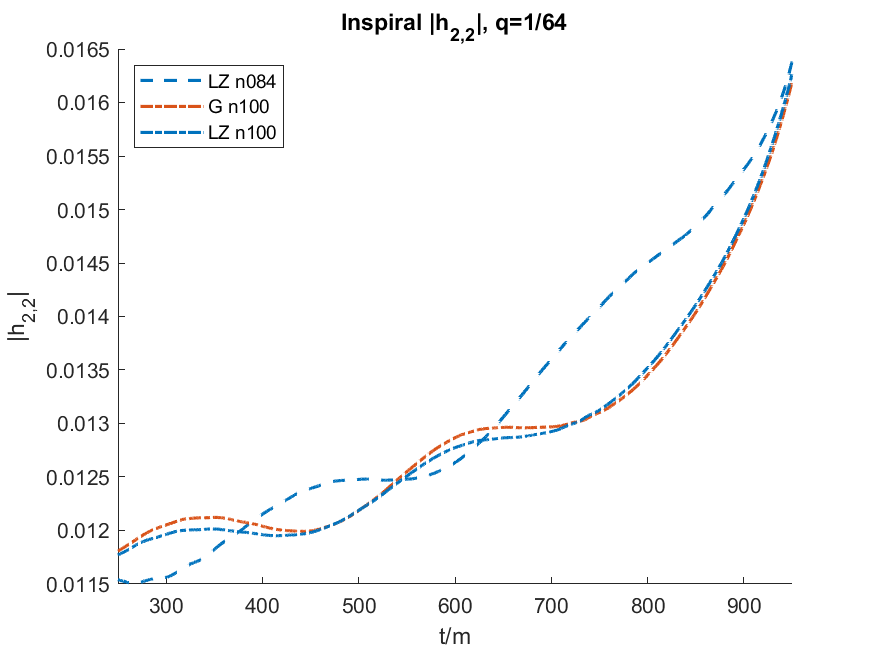}\\
\includegraphics[angle=0,width=0.99\columnwidth]{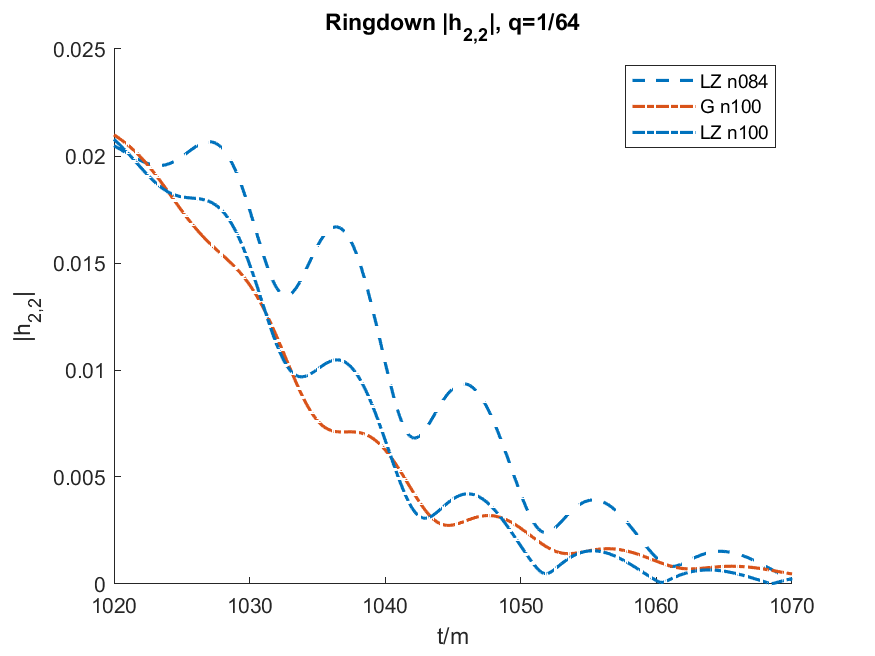}
\caption{Inspiral (top) and ringdown (bottom) amplitudes for the $q=1/64$ binary using the G and LZ gauges with resolutions n084 and n100. There is no clear difference between gauges in the inspiral, however, in the ringdown period, $h_{2,2}$ using the LZ gauge exhibits oscillations that are larger in amplitude and slightly higher in frequency than $h_{2,2}$ using the G gauge.}
\label{fig:Mergeq64}
\end{figure}

As the mass-ratio of the binary decreases, the inspiral and ringdown oscillations in the dominant mode of the gravitational wave strain increase in amplitude. Consider Fig.~\ref{fig:Mergeq64}. The top panel shows the inspiral, $t=(300-1300)m$ of $h_{2,2}$ for $q=1/64$ using gauges G and LZ in resolutions n084 and n100. Both G and LZ produce low-frequency oscillations in the inspiral. As shown in in the dot-dashed curves, these oscillations are comparable, but small, in amplitude at the production resolution n100. This is consistent with our findings for the $q=1/32$ binary, and is further evidence of the benefit of using a variable gauge during inspiral.

The bottom panel of Fig.~\ref{fig:Mergeq64} shows the ringdown phase of the evolution from $t=(1500-1530)m$. The oscillations in the ringdown of the simulation with the lowest resolution in the LZ gauge are fairly severe, however they damp away with an increase in resolution to n100.  With the use of the new G gauge, the oscillations in the ringdown of $h_{2,2}$ are decreased in both frequency and amplitude when compared to the corresponding ringdown in $h_{2,2}$ using the LZ gauge. We expect further improvements could be made with a gauge that has a near constant value post-merger, or an increase in resolution up to n144.

For such computationally demanding simulations as $q=1/64$ and $q=1/128$, the ability to accurately measure horizon quantities at lower resolution has substantial benefits in terms of computational expense. Table~\ref{tab:kickq64q128} shows the energy and angular momentum as measured on the horizon (columns 3 and 4), as well as the same quantities calculated from the amount of each carried away by gravitational waves (columns 5 and 6). It also includes the kicks for these simulations, but only measured by the amount of linear momentum carried away by gravitational waves (column 7). This measurement of linear momentum is significantly more consistent than using the measurement on the horizon, so those results are not included since they are not the main focus of this work. An improvement in the horizon measure of linear momentum would require an increase in resolution to n120 or even n144 which is not necessarily practical for such a small mass-ratio binary.

 This is consistent with our findings in for the $q=1/32$ binary, where the use of G and LZ to compute horizon linear momentum does improve with resolution but requires higher resolution simulations than those performed here.  This can be combined with our findings in~\cite{Healy:2020iuc}, in which a low-value constant $\eta$, such as 1 or 0.5, was able to very accurately measure recoil velocity on the horizon even at low resolutions, to construct a variable gauge that damps to a constant $\eta$ post-merger. 
 
 However, the other horizon quantities (mass and angular momentum) are also measured very accurately in both the G and LZ gauge. The exception, of course, is when resolution is too low (n084) to properly resolve the grid. There is an obvious improvement with respect to resolution, and an increase to n120 would likely produce horizon quantities that are on par with their radiated counterparts. 

\begin{table*}[t]
\begin{tabular}{|c|cc|cc|ccc|}\hline
\toprule
     &&&\multicolumn{2}{|c|}{Horizon}& \multicolumn{3}{|c|}{Radiated}\\
     q&Gauge&Resolution&$E/m$&$J/m^2$&$E/m$&$J/m^2$&$L$(km/s)\\\hline\hline
     1/64&G&n100&0.9990&0.0529&0.9990&0.0516&	2.8065\\\hline
     1/64&LZ&n084&0.9990&0.0457&0.9990&0.0516&	2.7216\\
     1/64&LZ&n100&0.9990&0.0519&0.9990&0.0516&	2.4303 \\\hline \hline
    1/128&LZ&n100&0.9996&0.0239&0.9995&0.0267&0.9703	\\\hline
     \bottomrule
\end{tabular}
\caption{The horizon quantities for the $q=1/64$ binary using the G and LZ gauges and the $q=1/128$ binary using the LZ gauge. The remnant quantities for the $q=1/128$ binary in the G gauge are not included since that simulation was not run through merger. The resolutions are given in column 3. The horizon quantities (energy and angular momentum) are in columns 4-5, and the corresponding radiated quantities are in columns 6-8. All simulations use CFL=1/4 and 8th order finite differencing stencils. The G gauge does a good job of measuring the horizon mass and spin, however, the horizon linear momentum requires high resolution to get close to the radiated measure, and is therefore not shown.}
\label{tab:kickq64q128}
\end{table*}


\subsection{Noise/spurious radiation versus mass ratio}\label{sec:NSvsq}

\begin{figure}[h!]
    \includegraphics[angle=0,width=0.99\columnwidth]{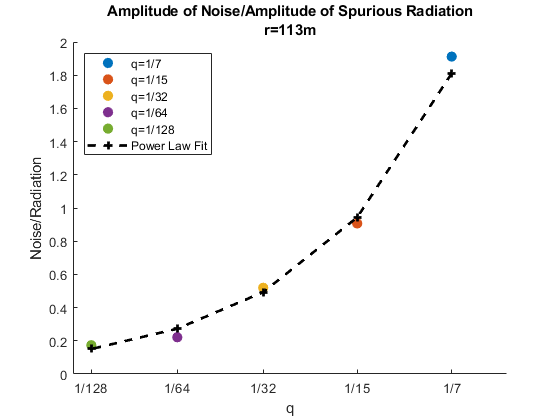}
    \caption{The data points show the ratio $\mathcal{N}_{peak}/\mathcal{S}_{peak}$ of $\Psi_4^{2,2}$ extracted at $r=113m$ for each mass ratio $q$. The black line is a power law fit to the data $F(q)=9.9753q^{0.8557}$. Errors are shown in Table~\ref{tab:noiseovrad}}.
    \label{fig:noiseovrad}
\end{figure}

The study of waveform amplitudes and phases can also provide an invariant measure of accuracy of the full numerical simulations. It has already been mentioned that this new gauge G completely damps the initial burst o noise at the beginning of the gravitational waveform, present when the LZ gauge is used, as well as its subsequent reflections at the refinement boundaries. In fact, this noise can be shown to be inversely proportional to the size of the peak in the spurious radiation generated at the beginning of the waveform due to initial data. Fig. \ref{fig:noiseovrad} shows the peak amplitude of the noise in $\Psi_4^{2,2}$, $\mathcal{N}_{peak}$, over the peak amplitude of the spurious radiation in $\Psi_4^{2,2}$, $\mathcal{S}_{peak}$, for each mass ratio $q=1/7,1/15,1/32,1/64,\text{ and }1/128$ extracted at $r=113m$. This figure uses the n100 simulations since we have those for every mass ratio.
We can generate a power-law fit that is a function of the mass-ratio $q$, which takes the form
\begin{equation}
F(q) = \mathcal{P}q^m
\end{equation}
where $m=0.8557\pm0.0448$ and $P=9.5739\pm0.0463$ are dimensionless constants. The fit has errors at each data point shown in Table \ref{tab:noiseovrad}.
\begin{table}[t]
    \begin{tabular}{|c|cc|}\hline
    $q$&$|F(q)-|\mathcal{N}_{peak}/\mathcal{S}_{peak}|$&$\dfrac{|F(q)-|\mathcal{N}/\mathcal{S}_{peak}|}{|\mathcal{N}_{peak}/\mathcal{S}_{peak}|}$\\\hline
    1/7&0.1012&0.0529\\
    1/15&0.0367&0.0405\\
    1/32&0.0260&0.0500\\
    1/64&0.0511&0.2308\\
    1/128&0.0229&0.1321\\\hline
    \end{tabular}
    \caption{The absolute and relative errors between $F(q)$ for all $q$ and $\mathcal{N}_{peak}/\mathcal{S}_{peak}$ and standard deviation for $\Psi_4^{2,2}$ extracted at $r=113m$ using resolution n100.}
    \label{tab:noiseovrad}
\end{table}
Although the absolute difference between the fit and data points scale with q, the relative error increases as q gets smaller. If we had higher resolution simulations for the smaller mass ratios, likely they would show that $\mathcal{N}_{peak}/\mathcal{S}_{peak}$ does in fact follow a power law. This means that the noise is in fact reduced significantly in amplitude in the small $q$-regime.

\section{Discussion and conclusions}\label{sec:discussion}

The gauge choices were crucial to the moving punctures breakthrough formalism that allowed numerical relativists to successfully evolve binary black holes \cite{Campanelli:2005dd,Baker:2005vv}. Here we explored choices of the initial lapse, $\alpha_0$, and shift, $\beta_0$, as well as the damping parameter $\eta$ in the equations (\ref{eq:gauge}).

In the construction of the waveform catalogs \cite{Healy:2017psd,Healy:2019jyf,Healy:2020vre} we used a Courant factor of 1/3 and sixth order spatial finite differencing for non-highly spinning and comparable masses binaries evolutions. This prioritized speed of the creation and therefore population of the initial catalogs to evaluate gravitational waves events observed by LIGO and Virgo \cite{Healy:2020jjs}. For highly spinning binaries \cite{Ruchlin:2014zva,Zlochower:2017bbg,Healy:2017vuz} and the small mass ratio systems we study here ($q\leq1/32$) require a reduction of the Courant factor to 1/4 and the use of 8th order finite differencing stencils to achieve good accuracy \cite{Zlochower:2012fk}. 

In our study of the improved choices of the initial lapse and shift inspired by their late time behavior we found some benefits in introducing the initial lapse as described in Section \ref{sec:7results}. In some preliminary studies we did on nonspinning binaries using an initial shift different from zero, we found the shift damped quickly to zero and then oscillated around zero during the binary's evolution before settling to the proposed initial form, without any obvious improvements to the system's physical parameters.

For comparable mass binaries of $q\geq 1/15$, the damping parameter $\eta$ is chosen to be a small, but constant value of order unity. Reduction of the mass ratio of the binary to $q=1/15$ and beyond benefits from $\eta$ variable in order to counteract the so called grid-stretching produced by the growth of the horizon in the numerical coordinates (See Fig.~\ref{fig:Ravg}) and thus depleting those gridpoints to resolve the fields in the exterior of the black holes. Of course, one can compensate by introducing more points via higher resolutions, but this comes at an increased computational cost.

The form of this damping parameter $\eta=$G given in Eq.~(\ref{eq:etaG}) provides a good general form valid for a wide range of mass ratios $q$. In addition to better maintaining of the physical parameters such as mass and spin, this choice of $\eta$ also removes unwanted initial noise in the waveforms, as well as its corresponding reflection on the mesh refinement levels when
compared the alternative original choice \cite{Lousto:2010ut,Lousto:2020tnb} (See also Eq.~(\ref{eq:eta0})) as displayed in figs.~\ref{fig:Psi4} and \ref{fig:Psi4q32}.
We have also found that the computational cost of introducing a variable $\eta$ is negligible compared to the whole evolution system of equations; the change is within 5\% of the speed of the $\eta2/m$ and is about 10\% faster than choosing LZ.

The constraint violation studies we performed are a useful measure of convergence with numerical resolution within a given gauge, but are not so useful when comparing different gauges.
We chose then to turn to analysis of physical parameters to determine the benefits of using different gauges.
The conservation of the horizon masses and spins (as shown in figs.~\ref{fig:mhor_7} and \ref{fig:mhor_15})
are a gauge independent measures of the accuracy of the simulations (Since absorption is an order of magnitude smaller of an effect during inspiral and merger).   
The evaluation of the horizon recoils also benefits from the property that G$\to1$ far from the black holes, as shown in Table~\ref{tab:kickq15}.

Additionally, we have looked in particular to the amplitude's behavior pre- and post- merger (see figs.~\ref{fig:Mergeq32} and~\ref{fig:Mergeq64}). To this end we have also introduced the $n=2$ in the $\eta_G$ choices in Eq.~(\ref{eq:etaG}) and fig. \ref{fig:etaGC} that smooths the values of $\eta$ around the smaller horizon, once the binary forms a common horizon (from when we normally drop the innermost refinement level).

In conclusion we recommend the use of the choices $(\alpha_0,\beta_0=0,\eta_G)$ parameters for the moving puncture evolutions due to its reasonable computational cost as well as the simplicity of its implementation. This gauge choice shows a wide range of improvements when dealing with essentially all possible mass ratios, $q\leq1$.




\begin{acknowledgments}
The authors thank Y.Zlochower for discussions on the gauge choices.
The authors gratefully acknowledge the National Science Foundation (NSF)
for financial support from Grants
No.\ PHY-1912632, No.\ PHY-1707946, No.\ ACI-1550436, No.\ AST-1516150,
No.\ ACI-1516125, No.\ PHY-1726215.
This work used the Extreme Science and Engineering
Discovery Environment (XSEDE) [allocation TG-PHY060027N], which is
supported by NSF grant No. ACI-1548562.
Computational resources were also provided by the NewHorizons,
BlueSky Clusters, and Green Prairies
at the Rochester Institute of Technology, which were
supported by NSF grants No.\ PHY-0722703, No.\ DMS-0820923, No.\
AST-1028087, No.\ PHY-1229173, and No.\ PHY-1726215.
Computational resources were also provided by the Blue Waters sustained-petascale computing NSF projects OAC-1811228, OAC-0832606, OAC-1238993, OAC- 1516247 and OAC-1515969, OAC-0725070 and by Frontera projects PHY-20010 and PHY-20007. Blue Waters is a joint effort of the University of Illinois at Urbana-Champaign and its National Center for Supercomputing Applications. Frontera is an NSF-funded petascale computing system at the Texas Advanced Computing Center (TACC).
\end{acknowledgments}

\bibliographystyle{apsrev4-1}
\bibliography{references}

\appendix

In this section we supplement the initial lapse and shift choices based on Lorentz boosted and Kerr spinning black holes.

\section{Initial lapse and shift for Boosted and Spinning Black Holes}\label{sec:LBKS}

\noindent {\it Lorentz-boosted black holes:}

Since we are considering binaries with initial orbital momentum, it is of interest to include corrections in the configurations of the initial lapse and shift that account for this. 

To construct initial lapse data for a black hole system with boost, the conformal factor $\psi_0$ is modified so that its order $\frac{1}{r}$ term is multiplied by the boost 
\[
\gamma = \frac{1}{\sqrt{1-|v|^2}}
\]
where $|v|$ is the magnitude of the boost velocity. The conformal factor becomes
\[
\psi_0 = 1+\frac{\gamma m}{2r}.
\]
It is then used in Eq.~(\ref{eq:LTL}) to calculate the initial boosted lapse.

To construct initial shift data for a black hole system with boost, we must calculate the shift term from a boosted Schwarzschild black hole metric in Cartesian coordinates. Then, $\beta^i$ terms can be added linearly onto the unboosted terms in Eq.~(\ref{beta0}).  

First, take the unboosted Schwarzschild metric 
\begin{equation}
g_{\mu\nu} = 
\begin{bmatrix}
-(\alpha)^2 &0&0& 0 \\
0&\psi_0^4 &0 & 0 \\
0 &0 & \psi_0^4& 0 \\ 
0 & 0& 0 &\psi_0^4 \\ 
\end{bmatrix}
\end{equation}
then apply a general boost transformation $\Lambda^\mu_{\nu}$ on the metric $g_{\mu\nu}$
\begin{equation}
\tilde{g}_{\mu\nu}  = \Lambda^\sigma_{ \mu} \Lambda^\xi_{ \nu} g_{\sigma \xi}
\end{equation} 
where
\begin{equation}
\Lambda_\mu^{\nu}=
\begin{bmatrix}
\gamma &-v_1\gamma&-v_2\gamma& -v_3\gamma \\
-v_1\gamma&1+ \frac{(\gamma-1)v_1^2}{|v|^2}& \frac{(\gamma-1)v_1v_2}{|v|^2} &  \frac{(\gamma-1)v_1v_3}{|v|^2} \\
-v_2\gamma & \frac{(\gamma-1)v_2v_1}{|v|^2}&1+  \frac{(\gamma-1)v_2^2}{|v|^2}&  \frac{(\gamma-1)v_2v_3}{|v|^2} \\ 
-v_3\gamma &  \frac{(\gamma-1)v_3v_1}{|v|^2}&  \frac{(\gamma-1)v_3v_2}{|v|^2} &1+ \frac{(\gamma-1)v_3^2}{|v|^2} \\ 
\end{bmatrix}.
\end{equation}
Then, the shift can be read off from the inverse spatial metric $\tilde{\gamma}^{ij}$ as  
\begin{equation}\label{betaboost}
\beta^i = \frac{v^i(-\eta^2 + \psi_0^2)}{|v|\eta^2-\psi_0^4}
\end{equation}
for $\psi_0(m,r) = 1+\frac{m}{2r}$, $\eta(m,r) = \frac{1-\frac{m}{2r}}{\psi_0} = \alpha_{sch}$, and
\[
v^i \approx \frac{P^i}{m}
\]
where $P^i$ are the momentum components of the boost. The terms $\beta^i$ are added linearly onto the corresponding terms in Eq.~(\ref{beta0}) to construct an analytic representation for a boosted, nonspinning BBH system.

\begin{figure}[H]
  \includegraphics[angle=0,width=0.8\columnwidth]{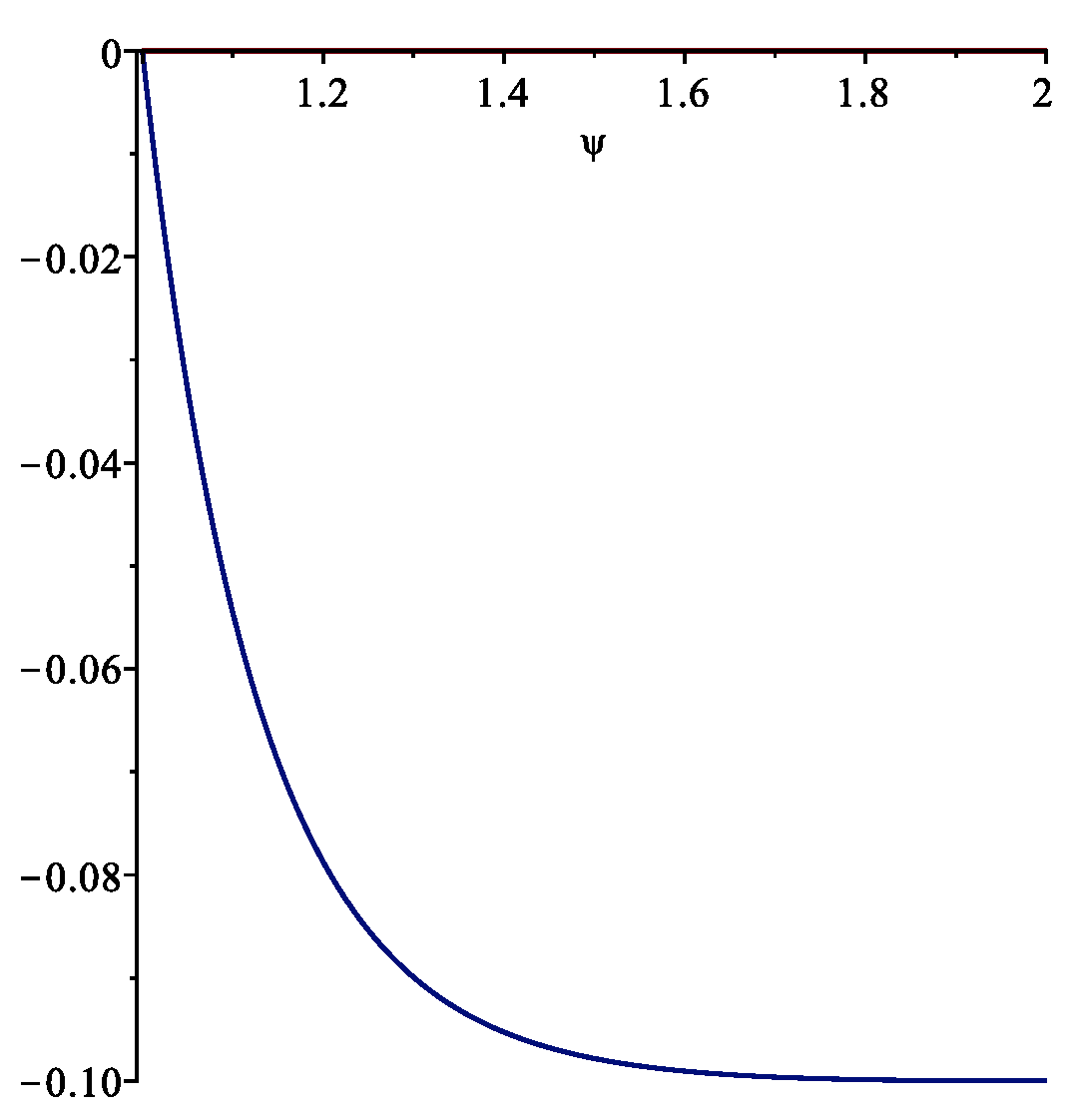}
  \caption{For small $r$, $\beta^y$ for an initial velocity $v=0.1$ versus 
$\psi_0=1+m/(2r)$. The horizon of this Schwarzschild black hole is located at $\psi_H=2$
and spatial infinity is at $\psi_\infty=1$. We see that the shift decays
slowly, like $\psi_0-1$.
    \label{fig:boostBy}}
\end{figure}

 We can try to force a stronger fall-off via asymptotic
matching or an attenuation function. 
Another alternative is to directly consider the trumpet Initial Data 
for Boosted Black Holes: See details in Ref.~\cite{Slinker:2018ujj}.
For trumpet Slices in Kerr Spacetimes, 
see details in Ref.~\cite{Dennison:2014eta}.

\bigskip

\noindent {\it Spinning black holes:}

To construct initial lapse data for a black hole system with spin, the conformal factor $\psi_0$ is modified so that its leading order term includes the magnitude of the spin $a=S_z/m^2$. 
\[
\psi_0 = 1+\frac{\gamma \sqrt{m^2 + a^2}}{2r}.
\]
It is then used in Eq.~(\ref{eq:LTL}) to calculate the initial boosted lapse. 

To construct a spinning initial model for the shift, as in the boosted case, 
the spin terms are added linearly onto the boosted initial model for the shift. 
They are calculated from the conformal Kerr metric in Cartesian spacetime 
\begin{equation}
g_{\mu\nu}=
\begin{bmatrix}
(\sigma - 1)r^2/\rho^2 & a \sigma y /\rho^2 & - a \sigma x/\rho^2 & 0 \\
a\sigma y/\rho^2 & 1 + a^2 h y^2 &- a^2 h x y & 0 \\
-a \sigma x /\rho^2 & -a^2 h x y & 1 + a^2 h x^2 & 0 \\ 
0 & 0& 0 &1 \\ 
\end{bmatrix}
\end{equation}
with coordinates $x^\mu = (t,x,y,z)$ and 
\begin{align}
r&= \sqrt{x^2 + y^2 + z^2}, \\
\bar{r} &= r\left(1 + \frac{m+a}{2r}\right)\left(1 + \frac{m-a}{2r}\right), \\
\rho^2 &= \bar{r}^2 + \left(\frac{az}{r}\right)^2,\\
\sigma &= \frac{2m\bar{r}}{\rho^2},\\
h &= \frac{1 + \sigma}{\rho^2}{r^2},
\end{align}
where $r$ is the quasi-isotropic radial coordinate, $\bar{r}$ is the Boyer-Lindquist radial coordinate, and the spin $a = S_z/m^2$. The shift can be read off as 
\begin{equation}
\beta_i = g_{0i}
\end{equation}
for $i$ spatial. In practice, 
\begin{equation}
\beta^i = \gamma^{ij}\beta_j
\end{equation}
is used. The shift components also must be rotated so that they are valid for arbitrary spin orientations, not just spins along the $z-axis$. To do this, we can do three rotations of the shift vector and then sum up the results:

\begin{align}\
z &\to z,\\
y &\to y,\\
x &\to x.\\
\end{align}
\begin{align}
z &\to x,\\
y &\to y,\\
x &\to -z.\\
\end{align}
\begin{align}
z &\to y,\\
y &\to -z,\\
x &\to x.\\
\end{align}
These rotations produce
\begin{widetext}
\begin{align}
\beta^x &= \frac{a_x a_z^2 x y z \rho_y^2 \sigma_x(1+\sigma_z) - \rho_x^2 ( -a_z r^2 y \rho_y^2\sigma_z + a_y z \sigma_y(r^2 \rho_z^2 + (a_z x)^2(1+\sigma_z))}{(\rho_x\rho_y)^2 ((r\rho_y)^2+a_z^2(x^2+y^2)(1+\sigma_z)}, \\
\beta^y &= \frac{a_x z \rho_y^2 \sigma_x(r^2\rho_z^2 + a_z^2 y^2(1+\sigma_z))-a_zx \rho_x^2 (r^2 \rho_y^2 \sigma_z + a_ya_z y z \sigma_y(1+\sigma_z))}{(\rho_x\rho_y)^2 ((r\rho_y)^2+a_z^2(x^2+y^2)(1+\sigma_z)}, \\
\beta^z &= \frac{- a_x y \sigma_x}{\rho_x^2} + \frac{a_y x \sigma_y}{\rho_y^2}
\end{align}
\end{widetext}
as the spin-corrected terms for the Kerr initial shift.

\end{document}